\documentclass[acmtog]{acmart}
\acmSubmissionID{1755}

\usepackage{booktabs} 

\usepackage[ruled]{algorithm2e} 

\usepackage{color}
\definecolor{bittersweet}{rgb}{1.0, 0.44, 0.37}
\definecolor{cblue}{rgb}{0.27, 0.76, 0.69}
\usepackage{booktabs} 
\usepackage{wrapfig}
\usepackage{overpic}
\usepackage{multirow}

\SetAlFnt{\small}
\SetAlCapFnt{\small}
\SetAlCapNameFnt{\small}
\SetAlCapHSkip{0pt}

\acmJournal{TOG}

\begin{document}
\title{NeuVAS: Neural Implicit Surfaces for Variational Shape Modeling}

\setcopyright{cc}
\setcctype{by-nc}
\acmJournal{TOG}
\acmYear{2025} \acmVolume{44} \acmNumber{6} \acmArticle{} \acmMonth{12} \acmPrice{}\acmDOI{10.1145/3763331}

\author{Pengfei Wang}
    \orcid{0000-0002-0938-267X}
    \affiliation{%
    \institution{Shandong University}
    \city{Jinan}
    \country{China}}
    \email{8144756@qq.com}

\author{Qiujie Dong}
    \orcid{0000-0001-6271-2546}
    \affiliation{%
    \institution{Shandong University}
    \city{Qingdao}
    \country{China}}
    \affiliation{%
    \institution{The University of Hong Kong}
    \city{Hong Kong}
    \country{China}}
    \affiliation{%
    \institution{TransGP}
    \city{Hong Kong}
    \country{China}}
    \email{qiujie.jay.dong@gmail.com}

\author{Fangtian Liang}
    \orcid{0009-0002-8399-6853}
    \affiliation{%
    \institution{Shandong University}
    \city{Jinan}
    \country{China}}
    \email{lft00@foxmail.com}
    
\author{Hao Pan}
    \orcid{0000-0003-3628-9777}
    \affiliation{%
    \institution{Tsinghua University}
    \city{Beijing}
    \country{China}}
    \email{haopan@tsinghua.edu.cn}
    
\author{Lei Yang}
    \orcid{0000-0002-3284-4019}
    \affiliation{%
    \institution{University of Hong Kong}
    \city{Hong Kong}
    \country{China}}
    \email{yanglei.dalian@gmail.com}

\author{Congyi Zhang}
    \orcid{0000-0002-4259-2863}
    \affiliation{%
    \institution{University of British Columbia}
    \city{Vancouver}
    \country{Canada}}
    \email{zhcy@outlook.com}
    
\author{Guying Lin}
    \orcid{0000-0002-5233-3910}
    \affiliation{%
    \institution{University of Hong Kong}
    \city{Hong Kong}
    \country{China}}
    \email{guyinglin2000@gmail.com}

\author{Caiming Zhang}
    \orcid{0000-0003-0217-1543}
    \affiliation{%
    \institution{Shandong University}
    \city{Jinan}
    \country{China}}
    \email{czhang@sdu.edu.cn}

\author{Yuanfeng Zhou}
    \orcid{0000-0001-6950-3261}
    \affiliation{%
    \institution{Shandong University}
    \city{Jinan}
    \country{China}}
    \email{yfzhou@sdu.edu.cn}
    
\author{Changhe Tu}
    \orcid{0000-0002-1231-3392}
    \affiliation{%
    \institution{Shandong University}
    \city{Qingdao}
    \country{China}}
    \email{chtu@sdu.edu.cn}
    
\author{Shiqing Xin}
    \orcid{0000-0001-8452-8723}
    \affiliation{%
    \institution{Shandong University}
    \city{Qingdao}
    \country{China}}
    \email{xinshiqing@sdu.edu.cn}

\author{Alla Sheffer}    
    \orcid{0000-0001-9251-3716}
    \affiliation{%
    \institution{University of British Columbia}
    \city{Vancouver}
    \country{Canada}}
    \email{sheffa@cs.ubc.ca}

\author{Xin Li}
    \orcid{0000-0002-0144-9489}
    \affiliation{%
    \institution{Texas A\&M University}
    \city{Texas}
    \country{United States of America}}
    \email{xinli@tamu.edu}

\author{Wenping Wang}
\authornote{Corresponding author: Wenping Wang.}     
    \orcid{0000-0002-2284-3952}
    \affiliation{%
    \institution{Texas A\&M University}
    \state{Texas}
    \country{United States of America}}
    \email{wenping@tamu.edu}

\begin{abstract}

Neural implicit shape representation has drawn significant attention in recent years due to its smoothness, differentiability, and topological flexibility.
However, directly modeling the shape of a neural implicit surface, especially as the zero-level set of a neural signed distance function (SDF), with sparse geometric control is still a challenging task.
Sparse input shape control typically includes 3D curve networks or, more generally, 3D curve sketches, which are unstructured and cannot be connected to form a curve network, and therefore more difficult to deal with.  While 3D curve networks or curve sketches provide intuitive shape control, their sparsity and varied topology pose challenges in generating high-quality surfaces to meet such curve constraints. 
In this paper, we propose NeuVAS, a variational approach to shape modeling using neural implicit surfaces constrained under sparse input shape control, including unstructured 3D curve sketches as well as connected 3D curve networks. 
Specifically, we introduce a smoothness term based on a functional of surface curvatures to minimize shape variation of the zero-level set surface of a neural SDF. We also develop a new technique to faithfully model $G^0$ sharp feature curves as specified in the input curve sketches.
Comprehensive comparisons with the state-of-the-art methods demonstrate the significant advantages of our method.

\end{abstract}


\begin{teaserfigure}
  \centering
  \vspace{-2mm}
  \includegraphics[width=\textwidth]{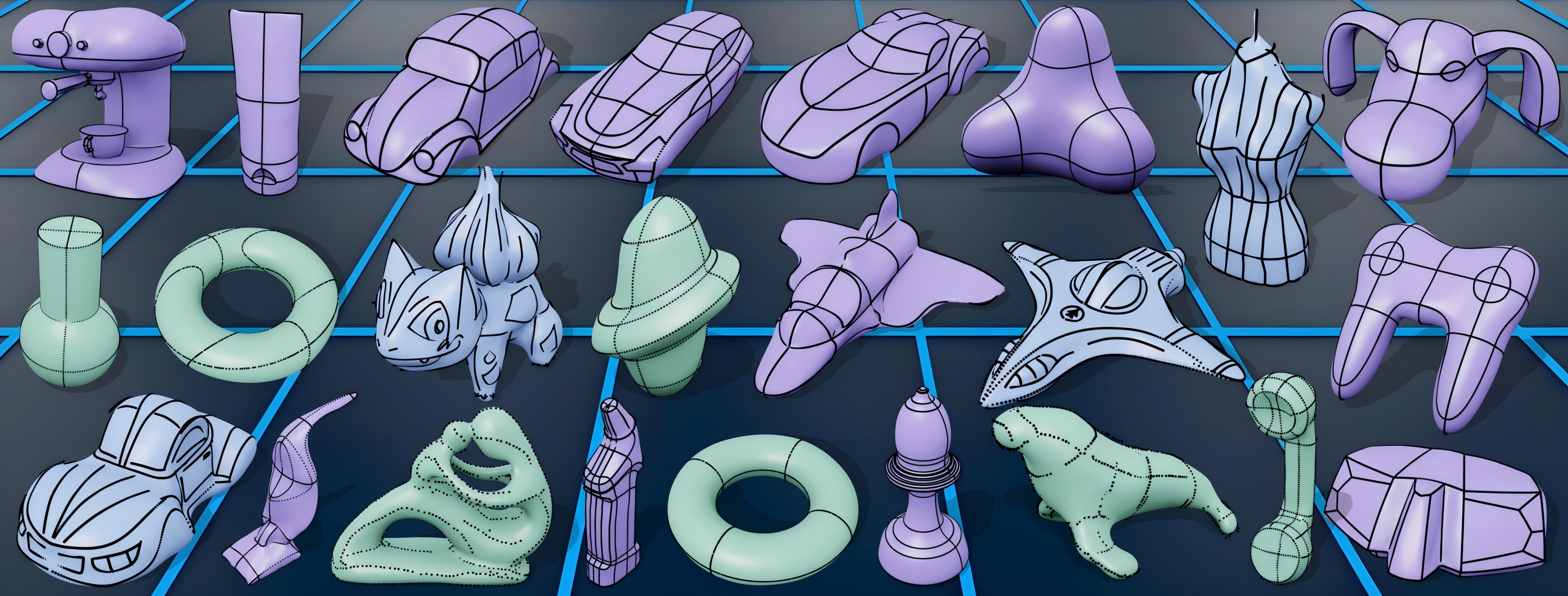}\\
  \vspace{2mm}
    \caption{A gallery of surfaces constructed using our method NeuVAS, with three kinds of sparse shape control input: connected curve networks (purple), unstructured curve sketches (blue), and sparse point clouds (green). 
    Connected curve networks are a set of curves connected to form loops to define surface patches.
    In contrast, unstructured curve sketches are a collection of disconnected curve segments defining the outline of a shape to be designed.
    Both curve networks and curve sketches can be discretized into point clouds, which serve as the input for NeuVAS.}
  \label{fig:teaser}
\end{teaserfigure}

\maketitle

\section{Introduction}
Surfacing a collection of 3D curves, also known as lofting or skinning, is a fundamental and challenging task in geometric modeling. This task involves three primary challenges: 1) finding a surface that accurately approximates the given curves; 2) controlling the shape away from the input curves to produce natural, aesthetic shape interpolation; and 3) constructing piecewise smooth surfaces to faithfully model $G^0$ sharp feature curves. In fact, most curve-based inputs are now manually drawn using AR/VR devices~\cite{yu2022piecewise}, and therefore often unconnected and inaccurate curve collections. This lack of structure makes the surface modeling even more difficult and 
renders some existing algorithms (e.g.~\cite{pan2015flow}) ineffective because they rely on structured curve networks as input.

 \begin{figure*}[!t]
  \begin{center}
  \includegraphics[width=0.95\textwidth]{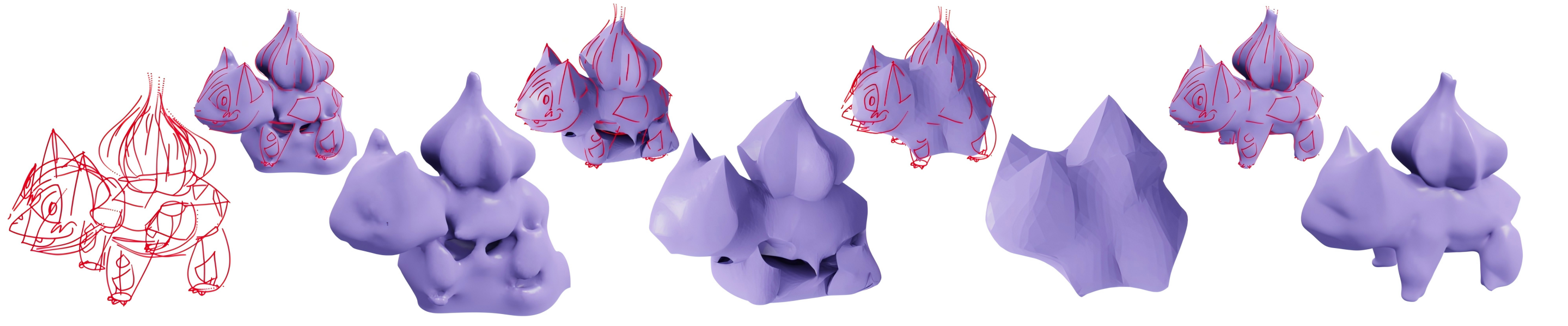}
  \makebox[0.19\textwidth][c]{\small (a) Input}
  \makebox[0.19\textwidth][c]{\small (b) VIPSS}
  \makebox[0.19\textwidth][c]{\small (c) Yu (+VIPSS)}
  \makebox[0.19\textwidth][c]{\small (d) Yu (+Sphere)}
  \makebox[0.19\textwidth][c]{\small (f) NeuVAS (Ours)}\\
  \makebox[0.19\textwidth][c]{\small }
  \makebox[0.19\textwidth][c]{\small ~\cite{huang2019variational}}
  \makebox[0.19\textwidth][c]{\small ~\cite{yu2022piecewise}}
  \makebox[0.19\textwidth][c]{\small ~\cite{yu2022piecewise}}
  \makebox[0.19\textwidth][c]{\small }
  \end{center}
   \caption{Comparisons with baseline methods capable of surfacing unstructured curve sketches as input. More comparison results are shown later in Section~\ref{subsec:comparisons}.}
   \label{fig:compare_intro}
\end{figure*}

Existing methods for surfacing 3D curve collections can be roughly divided into two categories: mesh-based methods~\cite{pan2015flow} and implicit surface methods~\cite{huang2019variational}.
The mesh-based methods can faithfully preserve sharp curve features by decomposing curve collections into independent loops to define individual patches. 
However, they require structured curve networks as inputs and cannot process unstructured curve sketches, thus limiting the scope of their application. 
Moreover, the performance of these methods is limited by the quality and resolution of mesh surfaces. 
Implicit surface methods, on the other hand, are more suitable for constructing surfaces from curve sketches due to their flexible topology and the access to accurate curvature computation. 
However, these methods face challenges in precise shape control, especially when constructing shapes with sharp features.

There is also a hybrid approach that combines both explicit and implicit surface representation~\cite{yu2022piecewise}.
The method of~\cite{yu2022piecewise} needs a mesh surface as input, which is presumably provided by another surfacing method from input curves. Then it segments the input mesh into patches and fits these patches with implicit polynomial surfaces.
Moreover, it needs a very good initial surface to start with to be successful. 
As acknowledged in~\cite{yu2022piecewise}, since the existing methods sometimes cannot produce such a good initial surface, one has to manually design an initial surface, making the method inefficient and unreliable. 
See Fig.~\ref{fig:compare_intro}, and more details of comparison in the Experiment Section.

\begin{figure}[ht]
  \begin{center}
  \includegraphics[width=0.9\columnwidth]{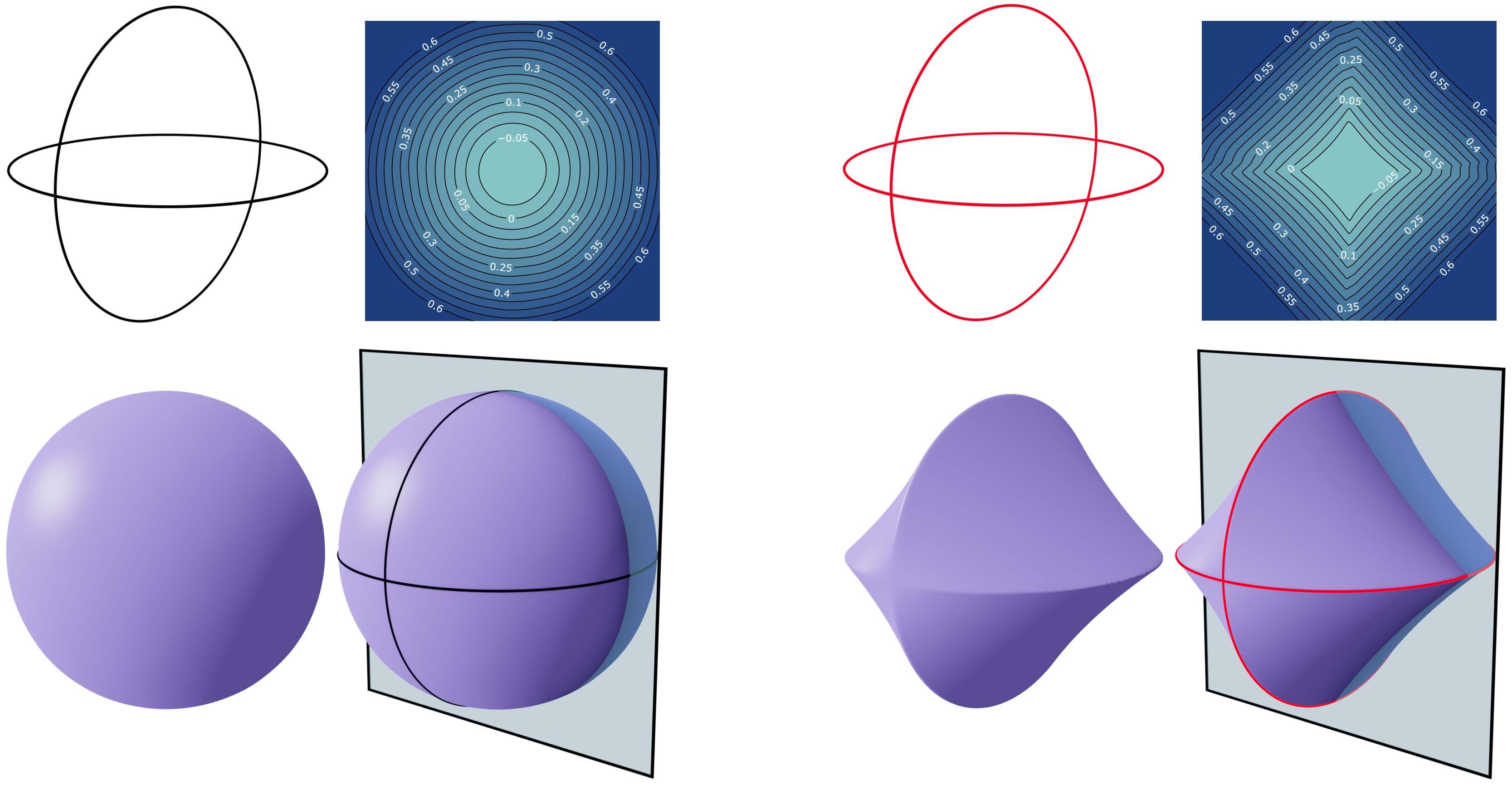}
  \makebox[0.45\columnwidth][c]{\small (a) Smooth}
  \makebox[0.45\columnwidth][c]{\small (b) Piecewise Smooth}
  \vspace{-4mm}
  \end{center}
   \caption{Shape control with sharp feature curves. (a) {\em The input curves contain no sharp features (in black)}: Using a smooth energy term, we produce a smooth transition between adjacent surface patches. (b) {\em The input curves are designated to be sharp features (in red) }: By reducing the influence of the smooth energy constraint near the input curves, sharp features are achieved.}
   \label{fig:field_2D_3D}
   \vspace{-2mm}
 \end{figure}

We propose a variational approach based on neural implicit surfaces for modeling high-quality surfaces from sparse shape control input, which including disconnected curve sketches, as well as connected curve networks and sparse points, therefore extending beyond only taking 3D connected curve networks as input. Here the neural implicit surface is modeled by the zero-level set of a neural Signed Distance Function (SDF) encoded by a Multi-Layer Perceptron (MLP). 
To regularize this surface, we use the Dirichlet condition~\cite{phase} and the Eikonal condition~\cite{IGR2020} to ensure interpolation of the curves. 
To regularize the surface shape in regions away from the input curves, we adopt a smoothness energy term based on the thin-plate energy~\cite{welch1994free} (Figure~\ref{fig:field_2D_3D}(a)).

To tackle the challenge of preserving sharp feature curves, we assign a location-dependent weight to the smooth energy term. 
More specifically, the weight at a surface location depends on its distance to sharp feature curves - the weight is larger when its location is away from the sharp feature curves, and the weight vanishes to zero when its location approaches the sharp feature curve. 
This strategy facilitates the modeling of shape feature curves because it separates the influences of the smoothness terms from the adjacent surfaces sharing a sharp feature curve, as shown in Figure~\ref{fig:field_2D_3D}(b). 
A gallery of surfaces constructed using NeuVAS from various sparse inputs is shown in Figure~\ref{fig:teaser}. 
Figure~\ref{fig:teaser} includes structured curve networks (purple) as well as more challenging inputs: unstructured and imperfect sketches (blue), and sparse points sampled along curves (green).
The latter two are commonly recognized as lower-quality input. 
The high-quality curves are typically used in CAD modeling, and the lower-quality inputs are more common in sketch-based modeling, especially in immersive VR/AR environments.
A key contribution of our method is to handle not only clean structured curves but also disconnected and inaccurate curve collections. To further validate the robustness of our method, we add Gaussian noises to the input curves and show the results in Figure~\ref{fig:abs_Noise}.

We note that the problem of variational shape modeling addressed in the present paper is fundamentally different from the conventional surface fitting task that seeks to over-fit a surface to a dense set of data points, possibly associated with surface normals. 
In the surface fitting task, the primary focus is the accurate approximation to the data points, therefore on-surface constraints and the Eikonal constraint for SDF regularization are sufficient.  
In contrast, for variational shape modeling, the input is a sparse set of 3D curves, which are insufficient to completely determine a definite surface shape. 
Therefore, in addition to ensuring interpolation of the curves, it is essential to include a surface energy term to regularize the surface shape in the region between the curves to produce an overall natural and aesthetic surface shape. 
In the Experiment Section~\ref{sec:compareSDF}, we will show that simply applying a surface fitting method to sparse input curves will lead to unsatisfactory results (Figure~\ref{fig:compare_SDFMethod}). 

Our main contributions are as follows:
\begin{itemize}
    \item We present a new method, NeuVAS, for high-quality variational surface modeling based on a neural implicit surface representation. Our method can take general sparse shape control as input, including disconnected curve sketches, as well as connected curve networks and sparse points. 
    
    \item We devise an effective strategy to preserve sharp feature curves on the output surface. This is achieved by properly reducing the influence of the surface energy constraints near designated sharp feature curves. 

    \item We conduct extensive experiments to validate our method and present comprehensive comparisons to demonstrate its advantages over existing methods. 
    
\end{itemize}

\section{Related Work}

\paragraph{Surfacing curve networks.}
The problem of surfacing curve networks emerged with the advent of practical user interfaces, algorithms, and devices designed to create these networks. 
Early 2D interfaces deduced the depth of curves from their intersections with other curves~\cite{schmidt2009analytic, xu2014true2form} or with sketching planes~\cite{bae2008ilovesketch}, effectively producing well-connected curve networks by design.
As a result, numerous surfacing algorithms heavily rely on the connectivity of the curve network to identify closed loops that delimit surface patches~\cite{abbasinejad2012surface, orbay2012sketch, sadri2014flow, zhuang2013general}.
Each such patch can then be surfaced by propagating geometric information from the boundary curves~\cite{zhuang2013general, pan2015flow, yu2021cassie}. 
Curve networks connectivity information provides strong geometric hints, as surface normals can be estimated at each intersection, as demonstrated by ~\cite{pan2015flow} to detect sharp features and determine which curves are flow lines. 
However, the reliance on accurate curve network topology prevents these methods from working on raw, unstructured stroke clouds typically produced via freehand AR/VR sketching.

Recent studies have shown that precise sketching in VR is more challenging than in 2D due to the lack of a supporting surface for the hand and the need for finer motor control to position curves in 3D~\cite{arora2017experimental, barrera2019effect}. Automatic snapping can help correct inaccuracies~\cite{machuca2018multiplanes, yu2021cassie}, but such assistance can be detrimental to user creativity. 
For instance, users of the recent CASSIE system~\cite{yu2021cassie} criticized the system for forcing them to think of their designs in terms of curve networks.
Processing 3D sketches to form well-connected curve networks is both challenging and often undesirable. 
First, many curves in 3D sketches are not intended to connect to others, and forcing such connections can alter the original design intent. Secondly, 3D sketches often exhibit oversketching and imprecision, making the formation of clean curve networks from 3D curves a challenging problem, similar to 2D vectorization.

We propose an approach that is oblivious to curve network connectivity, achieving high robustness to inaccuracies and to detailed curves that lie on the envisioned surface but are not connected to other curves.
Our work also differs from interactive systems designed to model 3D surfaces by drawing in 2D~\cite{dvorovzvnak2018seamless, li2017bendsketch, nealen2007fibermesh}. Users of these systems draw in dedicated interfaces and provide annotations of surface discontinuities. 
In contrast, our method takes point clouds as input and employs a smooth distance function to accurately recover sharp features.

\paragraph{Surfacing curve sketches.}
Curve sketches are typically manually drawn using AR/VR devices~\cite{rosales2019surfacebrush}. 
However, there are currently few methods that specifically target curve sketches. 
In the computer vision community, a few methods have been proposed to construct curve sketches by matching edges in multi-view stereo algorithms~\cite{fabbri20103d}. \cite{usumezbas2017surfacing} proposed a surfacing algorithm dedicated to such unstructured 3D curve sketches, where candidate surface patches are lofted between pairs of curves. 
However, this method heavily relies on the availability of multiple photographs of the shape to select valid candidate patches based on occlusion reasoning. \cite{batuhan2012free} surface sparse and imprecise curve sketches by smoothly deforming an initial low-fidelity surface of correct topology, using a discrete guidance vector field that points towards the closest curve point. 
This approach produces globally smooth surfaces and requires user intervention to specify curves that should be inserted into the mesh as sharp-edge polylines.

Recently, \cite{yu2022piecewise} introduced a multi-model fitting approach that generates piecewise-smooth surfaces. 
This method segments the surface into multiple patches, representing each patch as the zero-level set of an implicit polynomial. 
However, their approach has two notable shortcomings.
Firstly, it heavily relies on the accuracy of the initial surface; when the initial surface deviates significantly from the real geometry, the method fails to restore the correct shape. 
Secondly, the representation ability of implicit polynomials of degree up to 4 is limited, leading to the inability to recover small geometric details.
In contrast, our method does not require any other initialization methods; we simply take point clouds as input and leverage a neural network with strong representation abilities to learn the signed distance function (SDF).

\paragraph{Surfacing point clouds.}
The curve networks and curve sketches we consider can be easily converted to point clouds by sampling points along each curve. 
This approach enables us to leverage the wealth of methods developed for surfacing such unstructured 3D data~\cite{berger2017survey}. 
However, most existing methods are designed to process dense point clouds acquired using 3D scanning technology~\cite{hoppe1992surface, kazhdan2006poisson}, and thus struggle when applied to 3D sketches, which provide a very sparse, non-uniform sampling of the envisioned surface. 
While some methods are robust to missing data, they typically only address relatively small holes compared to the overall surface scale~\cite{hornung2006robust} or guide surface completion by fitting geometric primitives on dense parts of the point cloud~\cite{schnabel2009completion, tagliasacchi2009curve}. 

The thin-plate RBF~\cite{turk2002modelling} and VIPSS~\cite{huang2019variational} algorithm achieves impressive resilience to sparsity and non-uniformity of point clouds, and has even been demonstrated on unstructured point clouds. 
However, this robustness is achieved through a global smoothness energy that may overlook sharp surface features. 
Moreover, to avoid degenerate solutions such as a constant zero interpolant, thin-plate RBF requires additional spatial constraints with prescribed signed values.
Additionally, both thin-plate RBF and VIPSS have a computational complexity of $O(n^3)$, which limits their practical application to point clouds of up to around 5k points.
~\cite{xu2023globally} proposed point cloud orientation techniques and applied them to construct unstructured 3D data.
They compute the normal vectors based on global consistency and then use Poisson reconstruction~\cite{kazhdan2013screened} to generate the mesh surface.
However, their approach is ill-suited for reconstructing curve networks.
On one hand, ~\cite{xu2023globally} relies on the global consistency of normal vectors to generate smooth surfaces, which makes it incapable of handling curve networks with sharp features. 
On the other hand, even with accurately specified normal vectors, ~\cite{xu2023globally} struggles to generate a reasonable shape due to significant gaps in regions far from the curve network.
Furthermore, ~\cite{xu2023globally} exhibits high sensitivity to both the number of points and the shape of the curve network. It is limited to point cloud scales below 10K, rendering it inadequate for representing complex geometric shapes.

Our approach, based on neural signed distance functions (SDF)~\cite{jones20063d}, is self-supervised, and its loss includes a thin-plate energy term to constrain the empty space between the curve network while remaining unconstrained near feature curves. 
We optimize the patches on both sides of the feature curve independently, naturally achieving piecewise smoothness.

\section{Method}
Our NeuVAS method takes sparse shape control as input, including disconnected curve sketches, connected curve networks and sparse points, produces a neural implicit surface that is a natural, visually aesthetic surface, while satisfying the input curve constraints. 
The design of NeuVAS has three key aspects. 
(1) It uses the Dirichlet condition~\cite{phase} and the Eikonal condition~\cite{IGR2020} to ensure the input curves are accurately interpolated;
(2) NeuVAS uses a surface smoothness term based on the thin-plate energy to produce smooth and natural transitions between individual curve; 
and (3) NeuVAS relaxes the smooth constraint near those curves that are designated as sharp feature curves to yield a final surface that faithfully preserves the sharp features. These three aspects will be elaborated in the following subsections. 

\subsection{Interpolating Input Curves}
The input to our method is assumed to be curve sketches, which are typically a set of sparse, unstructured 3D curve segments. Given such curve sketches as input, our first goal is to compute a neural implicit surface that interpolates the individual input curves. This surface, denoted by $S$, is represented as the zero-level set surface of a neural Signed Distance Function (SDF), denoted as $f(\boldsymbol{x};\Theta):\mathbb{R}^3\to\mathbb{R}$,
which is encoded as an MLP, where $\Theta$ are the network parameters (i.e., weights). That is, $S=\{ \boldsymbol{x} \in {\mathbb R}^3 |   f(\boldsymbol{x};\Theta)=0\}$. 

Fitting a neural implicit surface $S$ to data points is a well-studied topic ~\cite{IGR2020, SIREN, SAL, NeuralPull, DiGS, park2019deepsdf}.
In the following, we provide a brief account of the loss terms used in our implementation, and refer to the details in the existing works, e.g.~\cite{SIREN}. For interpolating the curve sketches, our loss terms are based on the Eikonal condition, the Dirichlet condition, and a regularization term to avoid extraneous zero-level set points.

\paragraph{Eikonal Condition}
We use the Eikonal loss to promote an approximate SDF field. First, it helps regularize the implicit field, leading to cleaner zero-level sets, as discussed in~\cite{IGR2020}. Second, our point sampling strategy relies on having an approximate SDF to ensure distances and gradients are reliable near the surface. Let $\mathcal{P}$ denote the set of sample points of all the input curve sketches, where the SDF is supposed to be zero. Let $\mathcal{Q}$ be a bounding region of the input data, which is assumed to be normalized to the range~$[-0.5, 0.5]^3$ by default. 
By the Eikonal condition, an SDF $f(\boldsymbol{x};\Theta)$ possesses a unit gradient, i.e. $\| {\nabla_{\boldsymbol{q}}} f\|=1$, at any  point $\boldsymbol{q} \in \mathcal{Q}$. To constrain our neural function $f(\boldsymbol{x};\Theta)$ to represent an SDF, we use the Eikonal condition to define the loss term:
\begin{equation}
    \mathcal{L}_E = \frac{1}{|\mathcal{Q}|} \sum_{\boldsymbol{q} \in \mathcal{Q}} \left| 1 - \|\nabla f(\boldsymbol{q}; \Theta)\| \right|.
    \label{eq:loss_function}
\end{equation}

\paragraph{Dirichlet Condition}
To ensure that the neural implicit surface $S$ passes through the input curve sketches, 
we require that for any point $\boldsymbol{p} \in \mathcal{P}$, we have $f(\boldsymbol{p};\Theta) = 0$ as much as possible. This leads to the loss term:
\begin{equation}
    \mathcal{L}_{DM} = \frac{1}{|\mathcal{P}|} \sum_{\boldsymbol{p} \in \mathcal{P}} |f(\boldsymbol{p}; \Theta)|.
    \label{eq:1}
\end{equation}
Meanwhile, we keep the function $f$ from being zero at points away from the curve sketches, using the constraint term~\cite{SIREN}:
\begin{equation}
    \mathcal{L}_{\text{DNM}} = \frac{1}{|\mathcal{Q}|} \sum_{\boldsymbol{q} \in \mathcal{Q}} \exp(-\alpha |f(\boldsymbol{q}; \Theta)|).
    \label{eq:dnm_loss}
\end{equation}
Put together, the total loss for interpolating the curve sketches is:
\begin{equation}
\mathcal{L}_\text{interp} = \lambda_\text{E}  \mathcal{L}_\text{E} + \lambda_\text{DM} \mathcal{L}_\text{DM} + \lambda_\text{DNM} \mathcal{L}_\text{DNM}.
\label{Eqn:interp}
\end{equation}
This ensures that the neural function $f$ properly represents an SDF that accurately passes through the input curves. 

\subsection{Surface Smoothness Energy}

To encourage the final surface $S$ to take a natural smooth shape in regions between the sparse input curves, we adopt the following smoothness energy term, based on the thin-plate energy, as an additional loss term:
\begin{equation}
    \mathcal{L}_{\text{Smooth}} = \frac{1}{|S|} \sum_{\boldsymbol{s} \in S} \left(\kappa_1^2(\boldsymbol{s}) + \kappa_2^2(\boldsymbol{s})\right),
    \label{eq:smoothness_loss}
\end{equation}
where $\kappa_1$ and $\kappa_2$ are the principal curvatures of the underlying surface $S$. 
Let $H(\boldsymbol{s})$ and $K(\boldsymbol{s})$ denote the mean curvature and the Gaussian curvature at the surface point $\boldsymbol{s}$, respectively. 
Since 
$$H(\boldsymbol{s}) = [\kappa_1(\boldsymbol{s}) + \kappa_2(\boldsymbol{s})]/2, \;\;\; K(\boldsymbol{s}) = \kappa_1(\boldsymbol{s})\kappa_2(\boldsymbol{s}),$$ 
the term $\mathcal{L}_\text{Smooth}$
can be rewritten as:
\begin{equation}
\mathcal{L}_{\text{Smooth}} = \frac{1}{|S|} \sum_{\boldsymbol{s} \in S} \left( 4H^2(\boldsymbol{s}) - 2K(\boldsymbol{s}) \right),
\label{eq:placeholder}
\end{equation}
where $H(\boldsymbol{s})$ and $K(\boldsymbol{s})$ can be computed from the Hessian matrix ${\mathbb H}_f(\boldsymbol{s})$ of $f$ as follows ~\cite{Spivak1979GaussianCurvature, knoblauch1913GaussianCurvature, Ron2005GaussianCurvature}:
\begin{align}
    H(\boldsymbol{s}) =&  \frac
            {\nabla f(\boldsymbol{s};\Theta)  {\mathbb H}_f(\boldsymbol{s})  \nabla f^T(\boldsymbol{s};\Theta) - \| \nabla f(\boldsymbol{s};\Theta) \|^2  Trace({\mathbb H}_f(\boldsymbol{s}) )}
            {2 \| \nabla f(\boldsymbol{s};\Theta) \|^3}.\\
    K(\boldsymbol{s}) =& - \frac
            {\left | \begin{matrix}
                {\mathbb H}_f(\boldsymbol{s}) & \nabla f^T(\boldsymbol{s};\Theta) \\
                \nabla f(\boldsymbol{s};\Theta) & 0  \\
                \end{matrix} \right | }
            {\| \nabla f(\boldsymbol{s};\Theta) \|^4}.
\end{align} 

\paragraph{Remark.} Note that we use general curvature formulas instead of simpler formulas specific to SDFs. The reason is that, although our network is trained to approximate an SDF via the Eikonal loss, the resulting field is only an approximate SDF and may have considerable deviations from a true distance field, especially during training. As our focus is on accurately estimating curvature near the zero-level set, we chose to use the general formulas for mean and Gaussian curvature that apply to implicit fields, rather than relying on properties specific to true SDFs.

\subsection{Creating Sharp Feature Curves}
\begin{figure}[ht]
  \begin{center}
  \includegraphics[width=0.9\columnwidth]{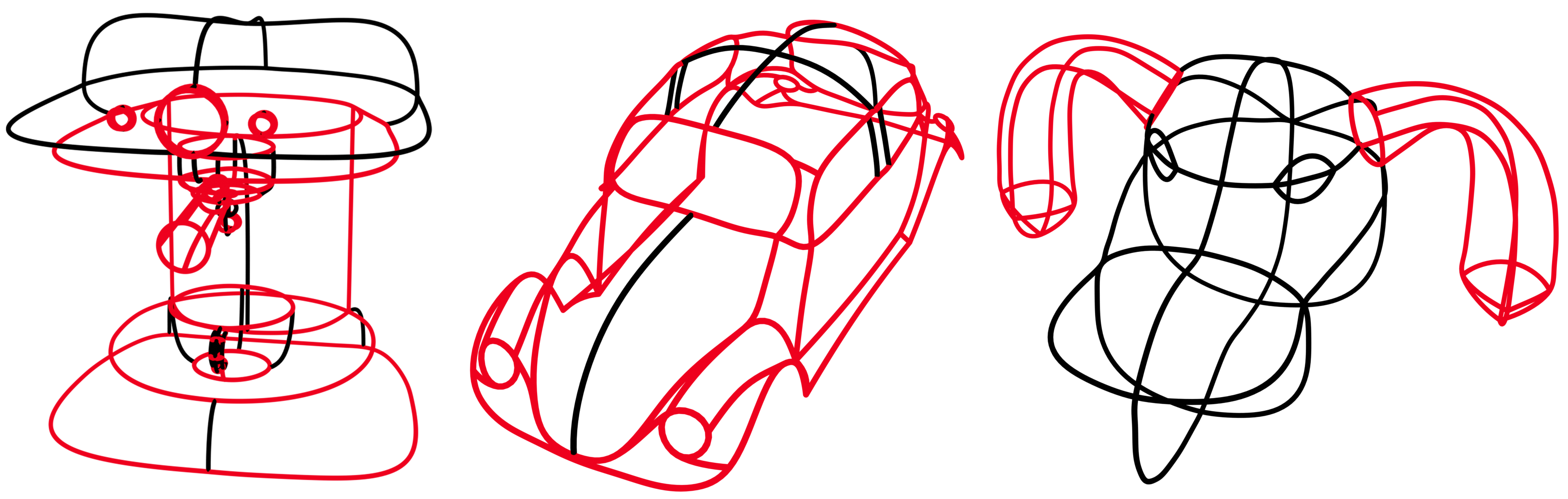}
    \makebox[0.27\columnwidth][c]{\small (a) Espresso}
    \makebox[0.27\columnwidth][c]{\small (b) Loftbug}
    \makebox[0.35\columnwidth][c]{\small (c) Doghead}
  \vspace{-4mm}
  \end{center}
   \caption{Results for detecting curve features in curve networks. The detected feature curves are shown in red. }
   \label{fig:feature_detect}
   \vspace{-1.4em}
 \end{figure}

In order to create a surface with sharp feature curves, i.e., curves along which the surface has $G^0$ continuity but non-continuous normal directions, the user may designate some of the input curves as feature curves. Let $\mathcal{P}$ denote the set of points on {\em all} the input curves (Fig~\ref{fig:feature_detect} both red and black curve), and $\mathcal{P}_f$ denote the set of points on the feature curves (Fig~\ref{fig:feature_detect} red curve), thus $\mathcal{P}_f \subset \mathcal{P}$.
To faithfully create the sharp features as specified, we need to reduce the influence of the smoothness term near and on the specified input feature curves; otherwise, enforcing the smoothness energy everywhere on the surface $S$ would result in an overall smooth surface without any sharp features, as shown in Fig.~\ref{fig:field_2D_3D}(a).

\begin{wrapfigure}{r}{0.2\textwidth}
\centerline{
    \includegraphics[width=0.95\linewidth]{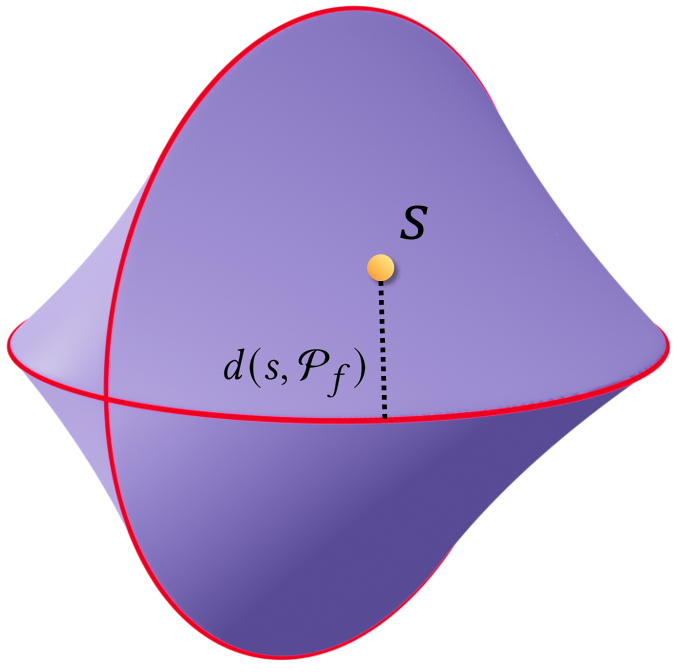}
    }
\end{wrapfigure}
To make the smoothness energy constraints decay around the feature curves, we introduce a weight function $d^2 (\boldmath{s}, \mathcal{P}_f)$, where $d(\boldmath{s}, \mathcal{P}_f)$ is the Euclidean distance from a surface point $\boldmath{s} \in S$ to $\mathcal{P}_f$, the set of points on the input feature curves. See the inset figure.
Then, incorporate this weight function to devise the following modified smoothness energy term:
\begin{equation} \label{eq:tps}
\mathcal{L}_{\text{Smooth}} = \frac{1}{|S|} \sum_{\boldsymbol{s} \in S} \left( 4H^2(\boldsymbol{s}) - 2K(\boldsymbol{s}) \right) \cdot d^2(\boldsymbol{s}, \mathcal{P}_f).
\end{equation}
The geometric intuition of this design is clear. The $\mathcal{L}_\text{Smooth}$ has small effect around the feature curves $P_f$ because the weight 
$d^2 (\boldsymbol{x}, \mathcal{P}_f)$ is small there, and $d^2 (\boldsymbol{x}, \mathcal{P}_f)=0$ for $\boldsymbol{x}$ on the feature curves. As a consequence, there is no smoothness constraint across the feature curves, leading to two adjacent surface patches sharing a feature curve to join with $G^0$ continuity. Furthermore, the absence of the smoothness energy would facilitate the accurate surface interpolation on the feature curve, where the loss term $\mathcal{L}_\text{interp}$ in Eq~\ref{Eqn:interp} is dominant. In our implementation, due to the normalization, the distance weight can be naturally bounded.

\paragraph{Designating Feature Curves}
For the curve networks, we extract feature curves by measuring the variation in surface normals at their two endpoints, as described in~\cite{pan2015flow}. These feature curves are detected automatically. For further details, please refer to~\cite{pan2015flow}.
Figures~\ref{fig:feature_detect} and~\ref{fig:ablation_curve_types} illustrate examples of detected feature curves and the resulting surfaces.
In the case of an curve sketches, where feature curves cannot be automatically extracted, we treat all curves as feature curves. For sparse point clouds, all curves are considered smooth curves by default.
Additionally, we support user-specified sharp feature curves.
In the following, feature curves are shown in red, and smooth curves in black.

\subsection{Implementation Details}\label{sec_Implementation}
Combining Eq.~\ref{Eqn:interp} and Eq.~\ref{eq:tps}, our total loss is given by:
\begin{equation}
\label{eq:our_loss}
    \mathcal{L} = \lambda_\text{E} \mathcal{L}_\text{E} + \lambda_\text{DM} \mathcal{L}_\text{DM} + \lambda_\text{DNM} \mathcal{L}_\text{DNM} + \tau\lambda_\text{Smooth} \mathcal{L}_\text{Smooth},
\end{equation}
where $\tau$ is the cosine factor~\cite{CosineFactor22} used to modulate the influence of the thin-plate energy loss term during training.
Empirically, the cosine factor $\tau$ is initialized to 1, with 1K iterations constituting one cycle.
Indeed, due to the conflict between the thin-plate energy term $\mathcal{L}_\text{Smooth}$ and the Dirichlet term $\mathcal{L}_\text{DM}$, the zero-level set of SDF $f$ struggles to interpolate the input curve network. 
We use the cosine factor to gradually reduce the thin-plate term $\mathcal{L}_\text{Smooth}$, encouraging the zero level set of SDF to interpolate the curve networks, and then increase the smoothing term $\mathcal{L}_\text{Smooth}$ to evolve towards a reasonable shape.
Note that once the surface aligns with the curve networks, the zero level set of SDF $f$ near the curve networks will not be constrained by the thin-plate term $\mathcal{L}_\text{Smooth}$ due to its distance based weighting, resulting in the surfacing sticking to the curve network. 
In practice, a complex model may contain features at significantly different scales; by leveraging the cosine factor with periodic annealing and restarts, our algorithm is able to recover these feature curves effectively.

\paragraph{Sampling on SDF Zero-Level Set}
To impose constraints on a surface region between curve, we need to sample at the zero-level set to measure the loss. 
However, accurate sampling at the zero-level set in each iteration would incur significant computational overhead, whether using Marching Cubes~\cite{lorensen1998marching} to extract zero-level set surfaces explicitly or iteratively projecting spatial samples away from the zero-level set onto it implicitly~\cite{RN590}. 
To mitigate this issue, we note that the change of the learned zero-level set between two consecutive iterations is small. 
Hence, for two iterations, we combine the strategy of zero-level set extraction and sample points projection. Namely, we reuse the point samples on the previous zero-level set and project them onto the current one.
Meanwhile, to ensure sampling uniformity and avoid cumulative errors, we schedule a step of Marching Cubes to regenerate sample points for every 100 iterations (including the initial iteration).
 
In particular, for a given neural SDF $f$, we extract the zero-level set mesh using Marching Cubes at resolution $128^3$. Then, we use Poisson-disk sampling on the zero-level set mesh to obtain uniform points. Through experiments, we find that setting the number of sampling points to 10K is reasonable and sufficient for almost all models, as shown in Figure~\ref{fig:teaser}.
In each iteration, we employ a projection strategy~\cite{RN590} to project the point set onto the zero-level. For a point $\boldsymbol{x}$, the projection $\boldsymbol{x}'$ is calculated as follows:
\begin{equation}\label{eq:project}
\boldsymbol{x}' = \boldsymbol{x} - \frac{\nabla f(\boldsymbol{x};\Theta)}{\|\nabla f(\boldsymbol{x};\Theta)\|} \cdot f(\boldsymbol{x};\Theta).
\end{equation}

\section{Experiments}\label{Sec:experiments}

\subsection{Experiment Setup}

\paragraph{Architecture.} 
Similar to various implicit surface reconstruction methods~\cite{IDF, DiGS, phase}, our NeuVAS employs the IGR~\cite{IGR2020} network architecture, featuring 8 hidden layers, each with 256 units, resulting in a model of 1.86M parameters. The activation function used is softplus. Inputs are normalized to the range $[-0.5, 0.5]^3$ before being fed into the network. 

\paragraph{Parameters.}

In our experiments, we utilized the thin-plate energy and determined the weights $\lambda_\text{E}=0.1$, $\lambda_{\text{DM}}=100$, and $\lambda_{\text{DNM}}=10$.
Empirically, we recommend setting $\lambda_{\text{Smooth}}$ to $5\times 10^{-4}$. In each iteration, we sample $\mathcal{Q}$ of 10K points uniformly inside the bounding box (normalized to $[-0.5, 0.5]^3$);
additionally, a $\mathcal{Q}_{zero}$ of 10K points are uniformly sampled on the zero-level set.
Throughout the training phase, we applied the Adam optimizer~\cite{Adam} with a default learning rate of $5 \times 10^{-5}$ and completed training in 10,000 iterations. For visualization, meshes are extracted from the zero-level set using the Marching Cubes algorithm at a consistent resolution of $512^3$ for all methods under comparison. The experiments were executed on an NVIDIA GeForce RTX 4090 graphics card equipped with 24GB of video memory along with an Intel(R) Core i9-13900k processor.

\paragraph{Datasets.}
The input data we use include curve networks~\cite{pan2015flow}, curve sketches~\cite{yu2022piecewise}, and sparse point clouds~\cite{huang2019variational}.

\begin{figure*}[ht]
  \begin{center}
  \includegraphics[width=0.95\textwidth]{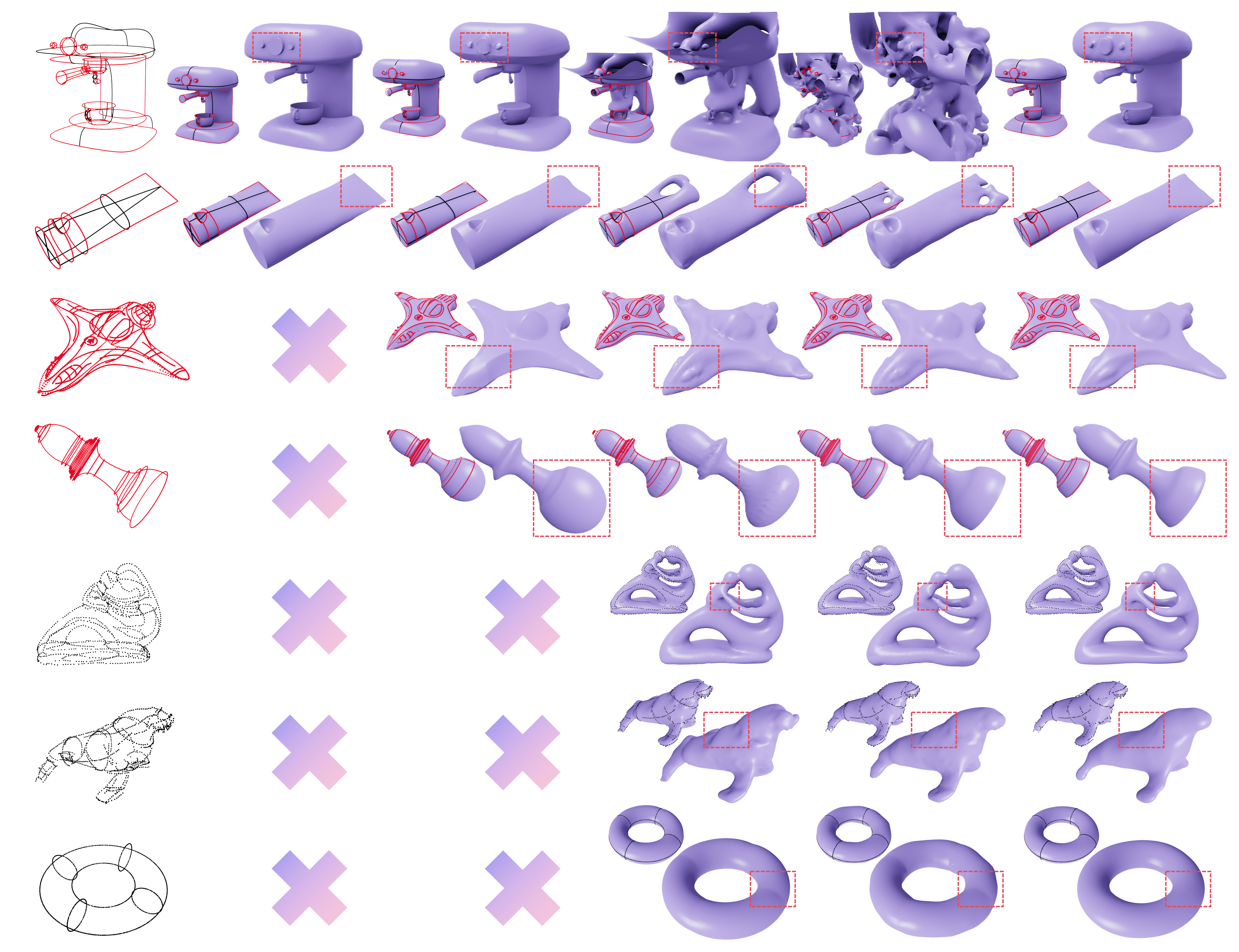}
    \makebox[0.155\textwidth][c]{\small (a) Input}
    \makebox[0.155\textwidth][c]{\small (b) ~\cite{pan2015flow}}
    \makebox[0.155\textwidth][c]{\small (c) ~\cite{yu2022piecewise}}
    \makebox[0.155\textwidth][c]{\small (d) ~\cite{huang2019variational}}
    \makebox[0.155\textwidth][c]{\small (e) ~\cite{xu2023globally}}
    \makebox[0.155\textwidth][c]{\small (f) Ours}
  \end{center}
   \caption{The comparisons on the dataset include three different input types: curve networks~\cite{pan2015flow}, curve sketches~\cite{yu2022piecewise}, and sparse point clouds~\cite{huang2019variational}. Note that, among the models produced by~\cite{yu2022piecewise}, two models Espresso and Toothpaste rely on their manually constructed initial proxy meshes.
   }
   \label{fig:all_comparisons}
\end{figure*}

\emph{Curve networks} are well-connected curves that can be decomposed into separate curve loops (e.g., Espresso and Toothpaste, the first and second rows in Figure~\ref{fig:all_comparisons}). However, in practice, the 3D sketch inputs typically do not have such clean structures.
\emph{Curve sketches} contain curves that cannot be clearly organized into loops and represent more frequent user creations through AR/VR devices (e.g., Spaceship and Bishop, the third and fourth rows in Figure~\ref{fig:all_comparisons}).
\emph{Sparse point clouds} contain scattered points that are hard to organize or connect into curves, making them unsuitable for methods like ~\cite{pan2015flow}, which requires closed loops, or ~\cite{yu2022piecewise}, which depends on explicit curves (e.g., Fertility, Walrus, and Torus, the fifth to seventh rows in Figure~\ref{fig:all_comparisons}).

\subsection{Results and Comparisons} \label{subsec:comparisons}
In this section, we compare our method with state-of-the-art variational surface modeling and neural implicit reconstruction methods. We then show extensive experiments evaluating runtime performance, the impact of different curve types, accuracy against ground-truth shapes, visualizations of energy distribution, and a discussion on how inputs affect the resulting shapes. We include more extensive experiments in the appendix and refer readers to our \emph{supplementary video} for a better visualization of our results using a turntable view.

\paragraph{Comparing to variational surface modeling methods} 
We compare our approach with 4 representative methods that are most relevant to variational surface modeling, including \cite{pan2015flow}, \cite{yu2022piecewise}, VIPSS~\cite{huang2019variational}, and ~\cite{xu2023globally}.

Among these methods, ~\cite{pan2015flow} is a mesh-based method that requires curve networks as input. With the high-quality curve network inputs, it can generate ideal shapes and recover sharp features in Espresso and Toothpaste (the first and the second row in Fig.\ref{fig:all_comparisons}).
However, due to the limitation of their curve loop-based algorithm, \cite{pan2015flow} cannot handle either curved sketches or sparse point clouds.

\cite{yu2022piecewise} uses a proxy mesh as input for initialization and then segments the mesh into different patches, hence it requires the proxy mesh to have the same topology as the expected result.
This requirement of the method for a very close proxy model makes it difficult to use. 
Specifically, such a proxy mesh needs to be generated using another method such as~\cite{huang2019variational} and sometimes the mesh thus generated does not work well; when this happens, the proxy mesh has to be constructed manually.
Therefore, for the models produced by~\cite{yu2022piecewise} that are included for comparison in Figure~\ref{fig:all_comparisons}, we generate their initial proxy mesh automatically using the output of~\cite{huang2019variational} as suggested by~\cite{yu2022piecewise}, and switch to manual construction whenever this automatic initialization fails. 
The models that need to rely on manual initialization are Espresso and Toothpaste in Figure~\ref{fig:all_comparisons}.
Note that \cite{yu2022piecewise} loses features at the bottom of Bishop and fails to capture fine geometric details at the front of Ship.

\cite{xu2023globally} first computes the normal vector of the point cloud, and then uses Poisson surface reconstruction~\cite{kazhdan2006poisson} to recover the shape.
Similarly, \cite{huang2019variational} models an implicit function that trades off between smoothness and fitting to sparse points.
Note that both ~\cite{xu2023globally} and ~\cite{huang2019variational} take only points as input and rely on the global consistency of normal vectors to generate smooth surfaces. As a result, they fail to achieve the correct topology when the input point clouds are too sparse (e.g., Espresso and Toothpaste), and they also cannot handle sharp features. Moreover, in the comparison, they introduce undesirable surface artifacts in Fertility and Walrus, such as unwanted bumps. In contrast, our method produces results that are more consistent with the ideal interpolated surfaces.
Besides, they are constrained by their algorithmic complexity ($\Omega(n^3)$) and limited to handling no more than 10K points.
However, a curve network with fewer than 10K points may not be sufficient to accurately represent complex models, e.g., the Espresso model shown in Figure~\ref{fig:all_comparisons} has 28K points.
To ensure proper execution, in this comparison, when the number of points exceeds 10K, we uniformly downsample them to 10K for both ~\cite{xu2023globally} and ~\cite{huang2019variational}.

\paragraph{Comparing to neural implicit reconstruction methods.}
\label{sec:compareSDF}
\begin{figure}[thb]
  \begin{center}
  \includegraphics[width=0.45\textwidth]{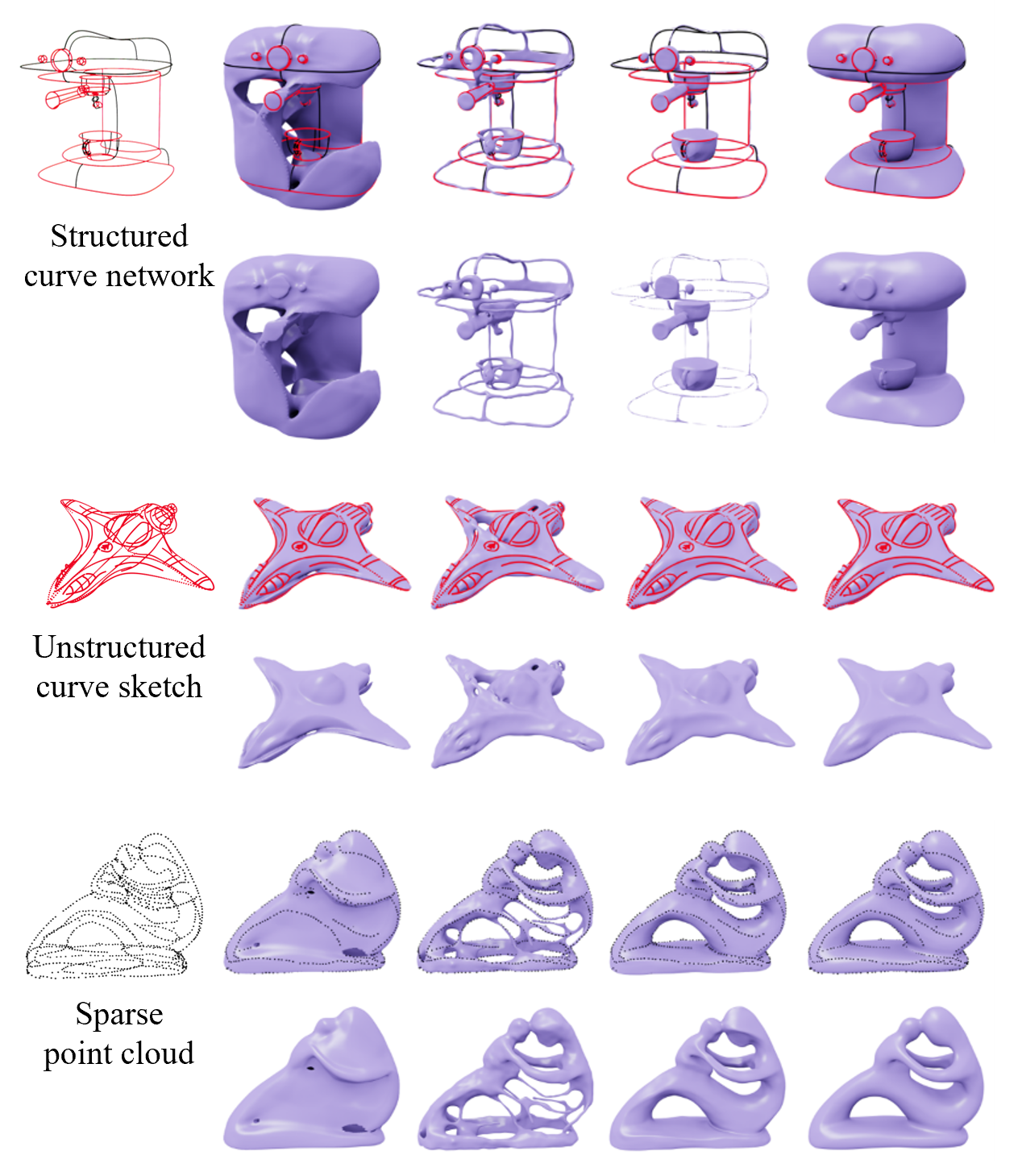}
    \makebox[0.18\columnwidth][c]{\small (a) Input}
    \makebox[0.18\columnwidth][c]{\small (b) IGR}
    \makebox[0.18\columnwidth][c]{\small (c) SIREN}
    \makebox[0.18\columnwidth][c]{\small (d) NSH}
    \makebox[0.18\columnwidth][c]{\small (e) Ours}
  \end{center}
   \caption{Comparison with surface reconstruction methods, including IGR~\cite{IGR2020}, SIREN~\cite{SIREN} and NSH~\cite{HessianZX}. These methods cannot yield visually pleasing shapes, due to the lack of surface modeling objectives away from the input points.}
   \label{fig:compare_SDFMethod}
\end{figure}
Due to the similarity in neural implicit representation, we have compared our method with the recent representative surface reconstruction works, such as IGR~\cite{IGR2020}, SIREN~\cite{SIREN}, and NeuralSingularHessian~\cite{HessianZX}. 
While these methods handle diverse input types, they cannot yield visually pleasing shapes, due to the lack of surface modeling objectives away from the input points.
As shown in Figure~\ref{fig:compare_SDFMethod}, our method produces much more plausible shapes than these methods.
All methods introduce incorrect topology or undesirable surface artifacts on our test shapes.
In contrast, our method reliably constructs the plausible geometry with fine details, correct topology, and sharp features.

\paragraph{Runtime performance.} 
\begin{table}[tb]%
  \centering 
  \caption{We report the time cost per iteration of our method across different zero-level set sample sizes, measured in milliseconds (ms). The mesh is first extracted and sampled, followed by the computation of the thin-plate energy for the samples using the neural network. }\label{tab:runtime_diff}
 \resizebox{\columnwidth}{!}{
   \begin{tabular}{c|c|c|c|c}
   \toprule
   Sample Size ($\mathcal{Q}_{zero}$)  &  10K  
                                       &  20K
                                       &  50K
                                       &  80K\\
   \hline
time~[ms]                            &  127.53
                                     &  132.39
                                     &  259.33
                                     &  387.13\\
   \bottomrule
 
 \end{tabular}
\centering
}
  \vspace{-2mm}
\end{table}

The computational cost primarily arises from evaluating the loss terms (Eq.~\ref{eq:our_loss}) on three distinct point sets: $\mathcal{P},\mathcal{Q},\mathcal{Q}_{zero}$.
Among these, the thin-plate energy estimation at $\mathcal{Q}_{zero}$ dominates the time complexity.
We fix the size of $\mathcal{P}, \mathcal{Q}$ to 10K, and reports the running time per iteration under different sample sizes for $\mathcal{Q}_{zero}$, as detailed in Table~\ref{tab:runtime_diff}.
Additionally, we apply MC to extract zero iso-surfaces once for every 100 iterations, so MC cost is amortized across the iterations. 
The results indicate that our method is still efficient as the number of sample points increases.
Thus in all experiments, we set the size of $\mathcal{P}, \mathcal{Q},\mathcal{Q}_{zero}$ to 10K and the MC resolution to $128^3$, resulting in a time cost of 127.53 ms per iteration.
The total number of training iterations is set to 10K, which is a conservative estimate; in most cases, convergence is achieved within 5K iterations.

\begin{table}[tb]%
  \centering 
  \caption{The run-time performance comparison, corresponding to the curve networks list in Figure~\ref{fig:all_comparisons} (top three rows), presents timing statistics in seconds (s).
  The time complexity of the method in ~\cite{huang2019variational} is $O(n^3)$, more than 10K points will time out, necessitating downsampling of the curve networks to 6K (Espresso), 4K (Toothpaste) and 4K (Roadster) vertices. 
  Similarly, the algorithm in ~\cite{xu2023globally} requires downsampling to 10K (Espresso), 5K (Toothpaste) and 5K (Roadster).}\label{tab:Time_compare}
 \resizebox{\columnwidth}{!}{
   \begin{tabular}{c|c|c|c}
   \toprule
   Method  &  Espresso  
           &  Toothpaste
           &  Roadster\\
   \hline
    ~\cite{pan2015flow}                  &  24.12
                                         &  20.73
                                         &  22.35\\
                                         
    ~\cite{yu2022piecewise}              &  527.84
                                         &  152.83
                                         &  51.37\\
   
    ~\cite{huang2019variational}         &  4274.32
                                         &  1321.65
                                         &  1274.53\\
                 
    ~\cite{xu2023globally}               &  1457.32 
                                         &  850.71
                                         &  328.47\\
                                         
    ~\cite{IGR2020}                      &  433.45 
                                         &  446.33
                                         &  442.12\\
                                         
    ~\cite{SIREN}                        &  113.94 
                                         &  124.75
                                         &  119.41\\
                                     
     ~\cite{HessianZX}                   &  411.22 
                                         &  408.31
                                         &  414.97\\
                                         
    Ours                                 &  1343.71
                                         &  1344.92
                                         &  1336.86\\
   \bottomrule
 
 \end{tabular}
\centering
}
  \vspace{-2mm}
\end{table}

The processing times for the curve networks in Figure~\ref{fig:all_comparisons} ((top three rows) are summarized in Table~\ref{tab:Time_compare}. 
The curve networks—Espresso, Toothpaste and Roadster—comprise 25K, 28K and 21K vertices, respectively. 
The method proposed by \cite{pan2015flow} can be decomposed into several stages: detecting closed loops\cite{zhuang2013general}, constructing the initial surface~\cite{zou2013algorithm,andrews2011linear}, and mesh optimization. 
Due to the decomposition of the curve network into multiple individual surface patches, this method lends itself to high parallelization. 
Although it offers the fastest running time, it is restricted to curve networks, significantly limiting its applicability.
The time complexity of \cite{huang2019variational} is $O(n^3)$, necessitating downsampling of the curve networks to 6K (Espresso), 4K (Toothpaste) and 4K (Roadster) vertices. 
Similarly, the algorithm in ~\cite{xu2023globally} is highly sensitive to both the number of points and the shape of the curve network, requiring downsampling to 10K (Espresso), 5K (Toothpaste) and 5K (Roadster). 
It suggests that both ~\cite{huang2019variational} and ~\cite{xu2023globally} lack the capability to handle complex models effectively.
The approach in ~\cite{yu2022piecewise} includes three stages: initial surface construction, iterative segmentation, and final mesh optimization. To mitigate the initialization effects, ~\cite{yu2022piecewise} requires multiple runs, selecting the solution with the lowest energy. 

In contrast, ~\cite{IGR2020}, ~\cite{SIREN}, ~\cite{HessianZX}, and our method leverage stochastic gradient descent (SGD) on input points, impose no such constraints on the number of input points. 
Although our method is slower than the other neural implicit reconstruction methods, they fail to generate plausible results for these inputs (Fig~\ref{fig:compare_SDFMethod}).

\paragraph{Effects of different curve types}

\begin{figure*}[!t]
  \begin{center}
  \includegraphics[width=0.95\textwidth]{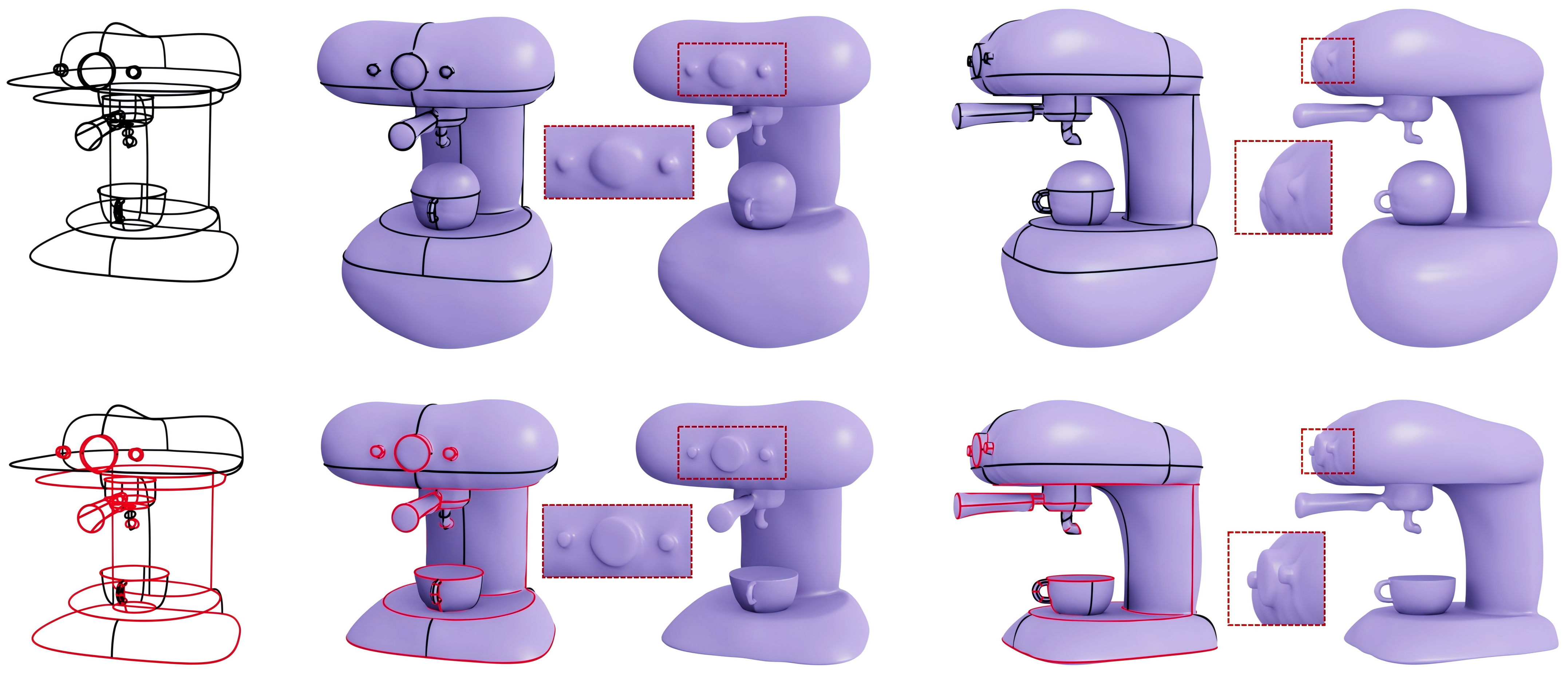}
    \makebox[0.19\textwidth][c]{\small (a) Input}
    \makebox[0.19\textwidth][c]{\small (b) Front view with curves overlaid}
    \makebox[0.19\textwidth][c]{\small (c) Front view}
    \makebox[0.19\textwidth][c]{\small (d) Side view with curves overlaid}
    \makebox[0.19\textwidth][c]{\small (e) Side view}
  \end{center}
   \caption{The effects  with or without marking the feature curves. Feature curves are highlighted in red. When the feature curves are well marked and represented in our NeuVAS, we generate both the sharp features and the smooth curves in the constructed model, resulting in a more reasonable shape, compared to setting all curves as smooth curves.
   }
   \label{fig:ablation_curve_types}
\end{figure*}

To justify our design choice of using two different types of curves, we show the effects of using the feature curve in Fig.~\ref{fig:ablation_curve_types}. Since this input is a curve network, we automatically detected feature curves by measuring the variation in surface normals at their two endpoints~\cite{pan2015flow}. The shapes are plausible when the feature curves are properly marked, and thus, the smoothness energy term does not dominate the shape control around the features. In comparison, when all curves are treated as smooth ones, it leads to an over-smoothed and bulged result (see the first row in Fig.~\ref{fig:ablation_curve_types}).

\paragraph{Evaluation against ground truth.}
\begin{figure}[htb]
  \begin{center}
  \includegraphics[width=0.45\textwidth]{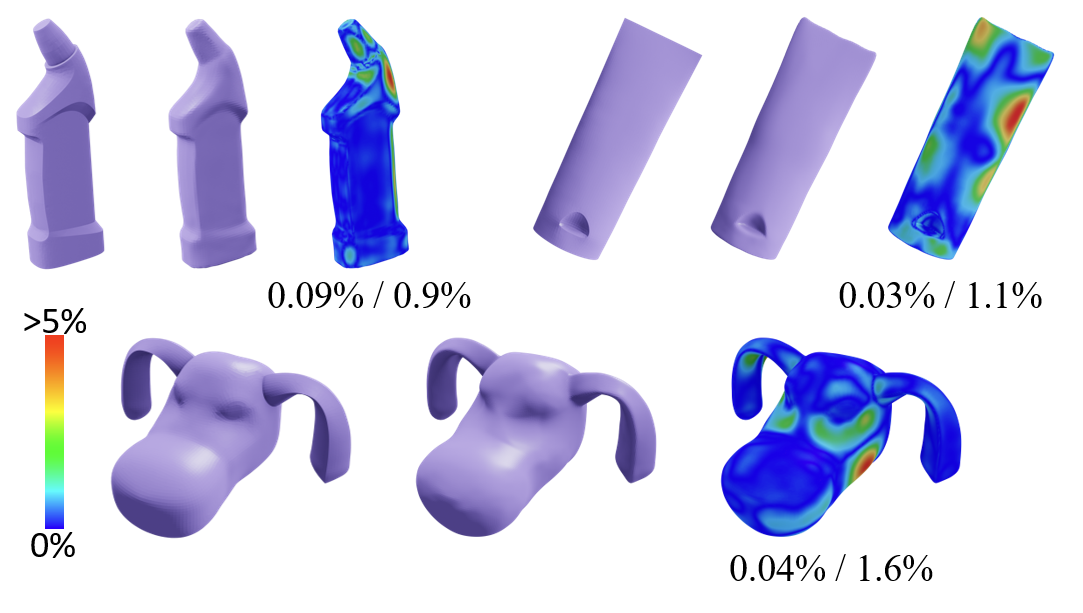}
  \makebox[0.33\columnwidth][c]{\small (a) GT}
   \makebox[0.3\columnwidth][c]{\small (b) Ours}
    \makebox[0.33\columnwidth][c]{\small (c) Error}
  \end{center}
   \caption{Comparing our results to ground truth, the color map shows the Hausdorff Distance from our result to Ground truth. The red region implies a deviation from the ground truth. The deviation is normalized by the bounding-box diagonal of the curve sketch. We provide the median and maximum deviation for each curve sketch.}
   \label{fig:CompareGT}
\end{figure}

We evaluate our method quantitatively by comparing our results to ground truth surfaces whose curve networks are available. In particular, we use the surface and curvature network data from FlowRep~\cite{FlowRep}, which generates descriptive curve networks from 3D shapes.
Figure~\ref{fig:CompareGT} visualizes the results of this experiment, where the color map represents the Hausdorff distance between our result and the ground truth shape.
The color map indicates that our results are very close to the ground truth.

\paragraph{Result energy distribution.}
\begin{figure}[tb]
  \begin{center}
  \includegraphics[width=0.45\textwidth]{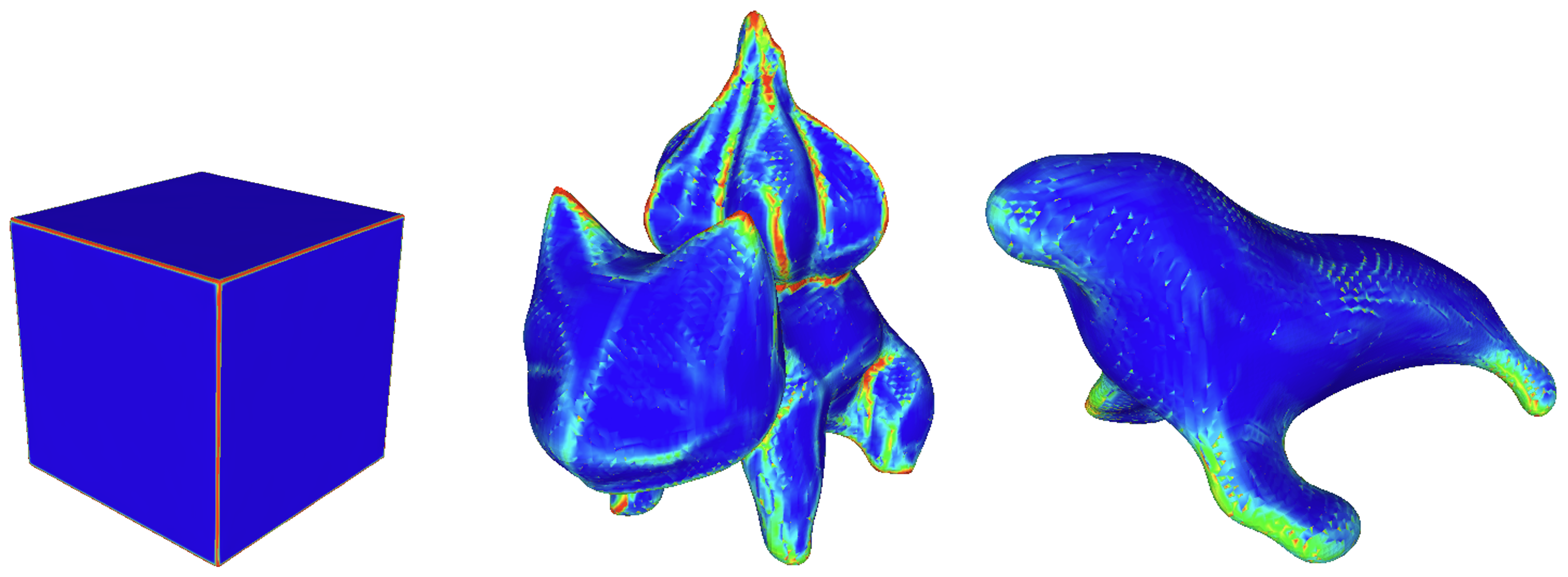}
    \makebox[0.3\columnwidth][c]{\small (a) Cube}
    \makebox[0.3\columnwidth][c]{\small (b) Bulbasaur}
    \makebox[0.3\columnwidth][c]{\small (c) Walrus}
  \end{center}
   \caption{Analysis of thin-plate energy distribution on results. Red color means a larger energy value. Although we use the distance attenuation strategy, the thin-plate energy in the empty region is evenly distributed.}
   \label{fig:abs_CurvatureMap}
\end{figure}
Figure~\ref{fig:abs_CurvatureMap} illustrates the spatial distribution of thin-plate energy across the surface. 
We colorize the thin-plate energy values with blue and red representing the minimum and maximum values, respectively. The color from blue to red corresponds to increasing thin-plate energy values. Notably, despite implementing a distance attenuation strategy, the thin-plate energy remains uniformly distributed in empty regions.

\begin{figure}[tb]
  \begin{center}
  \includegraphics[width=0.43\textwidth]{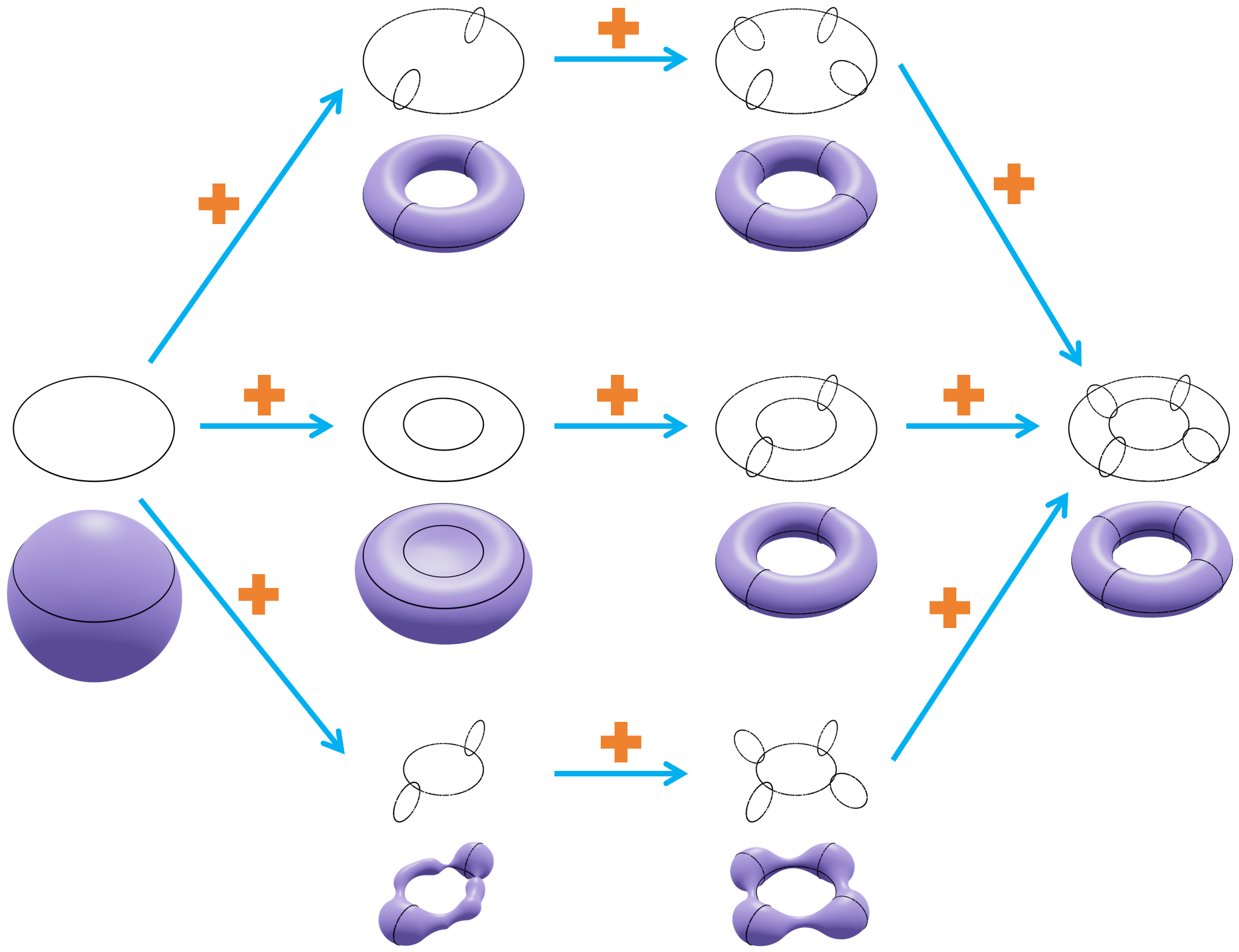}
    \makebox[0.24\columnwidth][c]{\small (a) }
    \makebox[0.24\columnwidth][c]{\small (b) }
    \makebox[0.24\columnwidth][c]{\small (c) }
    \makebox[0.24\columnwidth][c]{\small (d) }
  \end{center}
   \caption{
   Topology transition by incrementally adding curves to the input.   
   }
   \label{fig:ToursAdd}
\end{figure}

\paragraph{Topology transition.}
We conducted a test on our algorithm to see how much input is needed to produce plausible results.
Using a torus model, we progressively removed curves until only a single circle remained, then incrementally reintroduced the curves to observe the resulting variations.
Results are shown in Figure~\ref{fig:ToursAdd}.
Given a single circle as input, only a sphere is generated, which is a reasonable result.
As more guiding curves are added along the three different paths, we recover the torus shape at different steps, showing the impact of different curves and their combinations.

\section{Limitation}

\begin{figure}[ht]
\begin{minipage}{0.6\linewidth}
\centering
  \includegraphics[width=\linewidth]{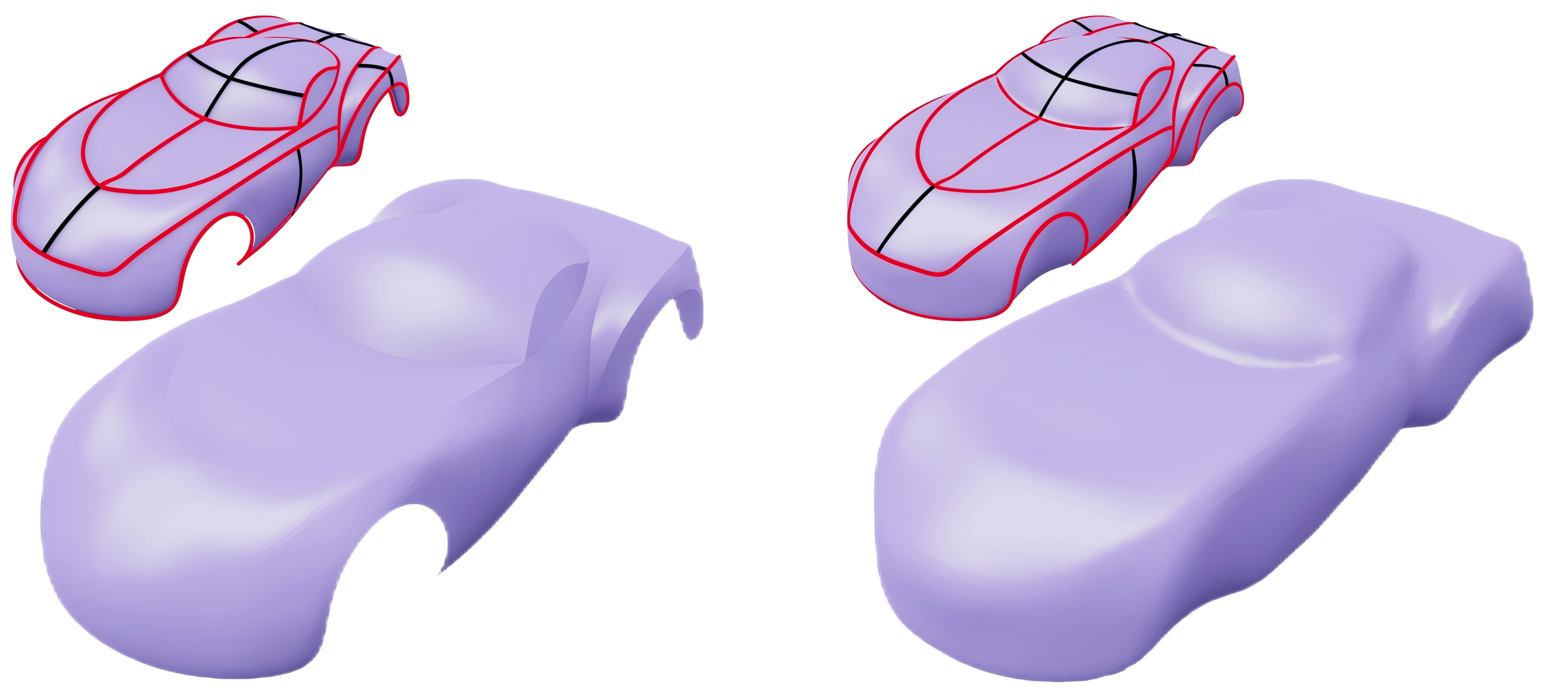}
  \makebox[0.48\linewidth][c]{\small (a) ~\cite{pan2015flow}}
  \makebox[0.48\linewidth][c]{\small (b) Ours}
\caption{Limitation of handling open surfaces.}
 \label{fig:opensurface}
\end{minipage}
\begin{minipage}{0.38\linewidth}
\centering
  \includegraphics[width=\linewidth]{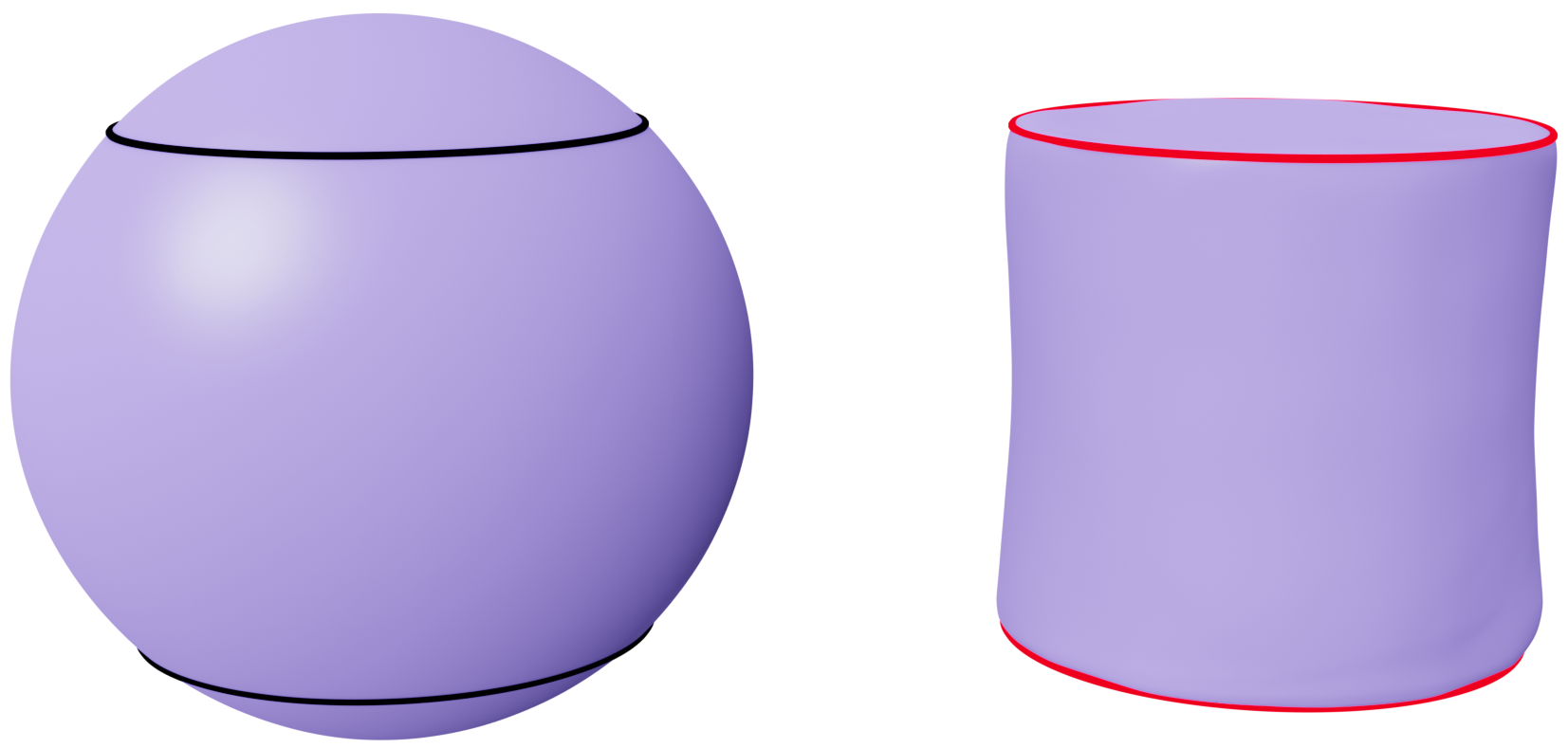}
   \caption{Imperfect spherical and cylindrical surfaces.}
   \label{fig:limit_cyl}
\end{minipage}
\end{figure}

\paragraph{Watertight manifold.}
NeuVAS is based on the assumption of a watertight manifold, which limits our ability to handle models with non-manifold geometry.
When dealing with open surface shapes, our method uses a watertight manifold as a prior to generate a closed surface, as shown in Figure~\ref{fig:opensurface}. 

\paragraph{Sharpness}
Due to the inherent limitations of the SDF representation, our method, despite achieving a high degree of sharpness, is unable to attain perfectly sharp features. This behavior is illustrated in Figure~\ref{fig:howsharpness}.

\paragraph{Euclidean distance weight}
In extreme cases, the shape may contain two very close, nearly parallel surfaces.
These surfaces are disconnected from each other, one is defined by a smooth curve, while the other is defined by a sharp curve.
Under such conditions, interactions between the two surfaces can occur, leading to slight irregularities or a reduction in smoothness.

\paragraph{Imperfect spherical or cylindrical shapes.}
In theory, the TPS energy converges to a state that tends towards local flatness, which excludes spherical or cylindrical shapes theoretically.
However, as shown in Figure~\ref{fig:limit_cyl} (a), in the presence of a smooth curve constraint, the TPS energy can still stabilize to a spherical shape whose residuals of two principal curvatures are evenly distributed (i.e. $k1 = k2$). 
On the other hand, the TPS energy does not stabilize on a cylinder, as shown in Figure~\ref{fig:limit_cyl} (b), making it slightly concave.
To this end, the curvature variation energy~\cite{Joshi2007CurvatureVariationEnergy,nealen2007fibermesh} permit to converge to cyclide shapes including spherical and cylindrical surfaces. 
However, being a third-order functional, it suffers from inherent instability and incurs substantial computational overhead.
We leave it as future work to address these challenges in our framework.

\begin{figure}[thb]
  \begin{center}
  \includegraphics[width=0.45\textwidth]{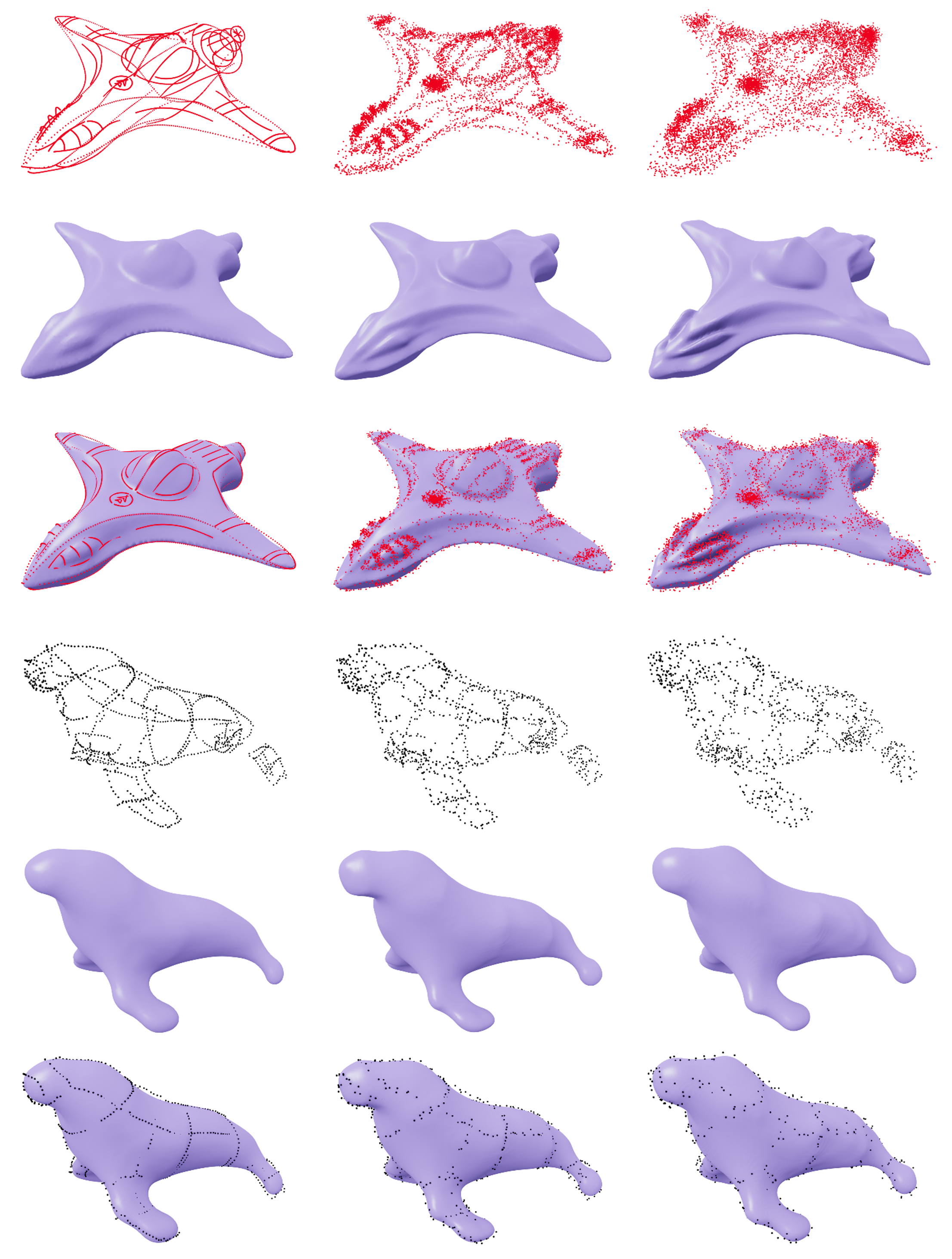}
  \makebox[0.3\columnwidth][c]{\small (a) No Noise}
  \makebox[0.3\columnwidth][c]{\small (b) $0.05$ Noise}
  \makebox[0.3\columnwidth][c]{\small (c) $0.1$ Noise}
  \end{center}
   \caption{Introducing Gaussian noise to the input curves. 
   NeuVAS successfully generates quality surfacing results at noise levels up to $0.05$. However, our method may not handle data with a noise level exceeding $0.1$.}
   \label{fig:abs_Noise}
\end{figure}

\paragraph{Noise.}
Our method demonstrates notable noise resistance, yet exhibits limitations when processing significantly noisy inputs, as illustrated in Figure~\ref{fig:abs_Noise}. In our experiments, Gaussian noise was introduced to the curve to evaluate robustness. The algorithm maintains its ability to generate high-quality surfaces at noise levels up to $0.05$. However, beyond a threshold of $0.1$ noise, the method fails to produce satisfactory results.

\section{Conclusion}
We introduce a variational surface modeling framework via representing signed distance function (SDF) encoded shapes by neural networks. 
Our framework incorporates the Dirichlet and Eikonal conditions to ensure boundary fidelity, while introducing a surface smoothness loss to regulate the shape in empty space between curves. 
This smoothness energy achieves piecewise smoothness, leaving regions near curves constrained by boundary conditions to preserve sharp features. 
Thanks to the neural implicit formulation, our method accepts point clouds as input, and flexibly changes surface topology to find plausible shapes with enhanced robustness. 
Through comprehensive comparisons with existing state-of-the-art methods, we demonstrate the significant advantages of our approach in constructing complex models.

\begin{acks}
The authors would like to thank the anonymous reviewers for their valuable comments and suggestions. This work was supported by the National Key Research and Development Program of china (2024YFB3309500), the National Natural Science Foundation of China (62402295, 62172257, U23A20312, 62272277), the Natural Science Foundation of Shandong Province (ZR2024QF087, ZR2024ZD12), the Innovation and Technology Commission of the HKSAR Government under the InnoHK initiative and the Hong Kong RGC (Ref. T45-205/21-N), the Joint Fund of the National Natural Science Foundation of China (U22A2033, U24A20219), the National Key R\&D Program of China (2022YFB3303200), and the Natural Sciences and Engineering Research Council of Canada (RGPIN-2024-03981).
\end{acks}

\bibliographystyle{ACM-Reference-Format}
\bibliography{sample-bibliography}

\appendix
\section{Additional Experiments}

We conducted extensive experiments to evaluate our method and analyze our results in the following aspects: loss functions, curve types, surface sampling numbers, convergence, initialization, result energy distribution, result topology change, and stress tests.

\subsection{Topological Change}
\begin{figure}[ht]
  \begin{center}
  \includegraphics[width=0.98\linewidth]{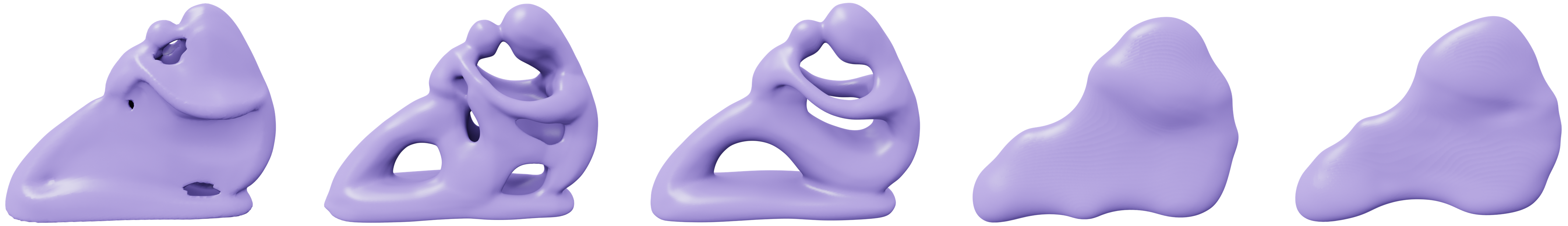}
  \makebox[0.185\columnwidth][c]{\small (a) $5 \times 10^{-7}$ }
    \makebox[0.185\columnwidth][c]{\small (b) $5 \times 10^{-6}$ }
    \makebox[0.185\columnwidth][c]{\small (c) $5 \times 10^{-5 \sim -4}$ }
    \makebox[0.185\columnwidth][c]{\small (d) $5 \times 10^{-3}$ }
    \makebox[0.185\columnwidth][c]{\small (e) $5 \times 10^{-2}$ }
  \end{center}
   \caption{The effect of thin-plate energy by varying its weight}
   \label{fig:Genus}
\end{figure}

\begin{figure}[ht]
  \begin{center}
  \includegraphics[width=0.98\linewidth]{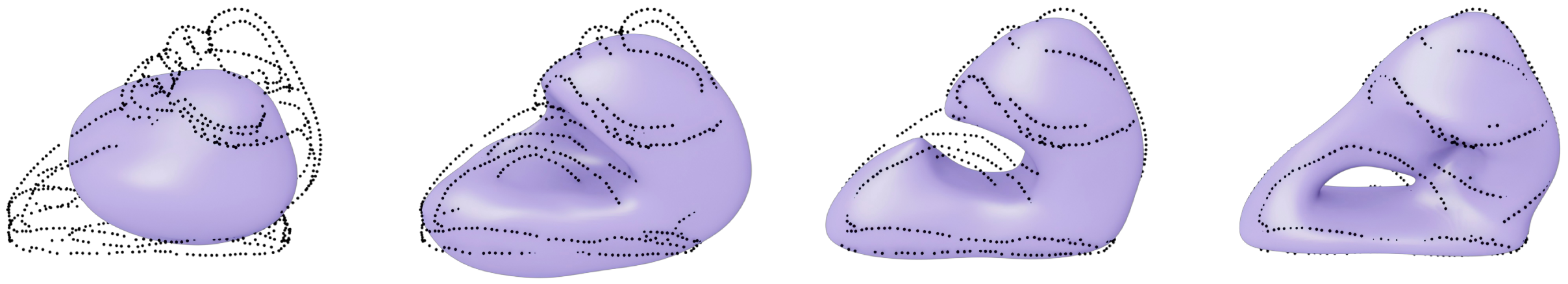}
    \makebox[0.24\columnwidth][c]{\small (a) $Iter:0.01K$}
    \makebox[0.24\columnwidth][c]{\small (b) $Iter:0.5K$}
    \makebox[0.24\columnwidth][c]{\small (c) $Iter:1K$}
    \makebox[0.24\columnwidth][c]{\small (d) $Iter:2K$}
  \includegraphics[width=0.45\textwidth]{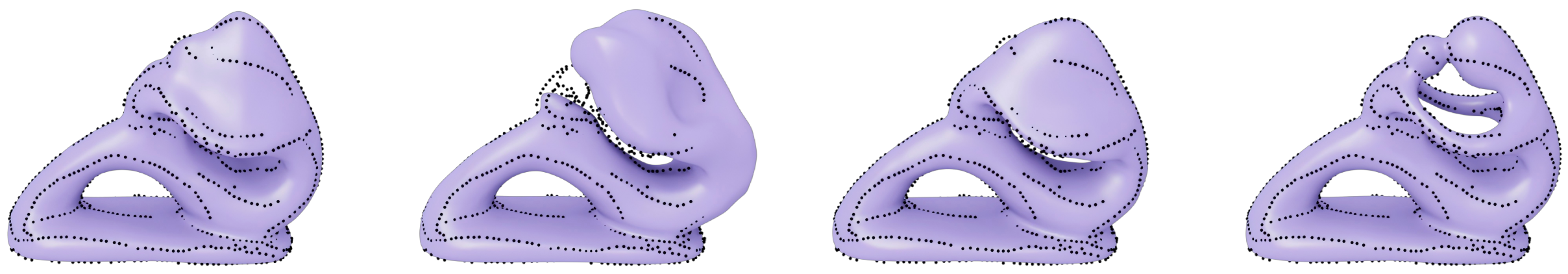}
    \makebox[0.24\columnwidth][c]{\small (e) $Iter:3K$}
    \makebox[0.24\columnwidth][c]{\small (f) $Iter:4K$}
    \makebox[0.24\columnwidth][c]{\small (g) $Iter:5K$}
    \makebox[0.24\columnwidth][c]{\small (h) $Iter:6K$}
  \end{center}
   \caption{Change of genus over iterations. During the evolution of shape, the surface remains smooth.} 
   \label{fig:Genusenvlop}
\end{figure}
We demonstrate the change of topology under different weights of the thin-plate energy. 
As show in Figure~\ref{fig:Genus}, when the weight is between $5 \times 10^{-5}$ and $5 \times 10^{-4}$, we achieve correct topology and obtain an ideal surface.
We also demonstrate the change of genus with iteration in Figure~\ref{fig:Genusenvlop}. 
As the iteration progresses, the genus continues to increase, and by the time the iteration reaches 6K, it has already converged.
Note that during the evolution of shape, the surface remains smooth throughout.

From these results, we can see that, while our method allows flexible change of topology, it is however, difficult to directly control the topology, for example by prescribing genus.
To this end, constraints in the line of \cite{ZouTopoConstrain15,HuangTopoControl17} may potentially allow for more direct control.

\subsection{Different Curve Types}
\begin{figure}[ht]
  \begin{center}
  \includegraphics[width=0.45\textwidth]{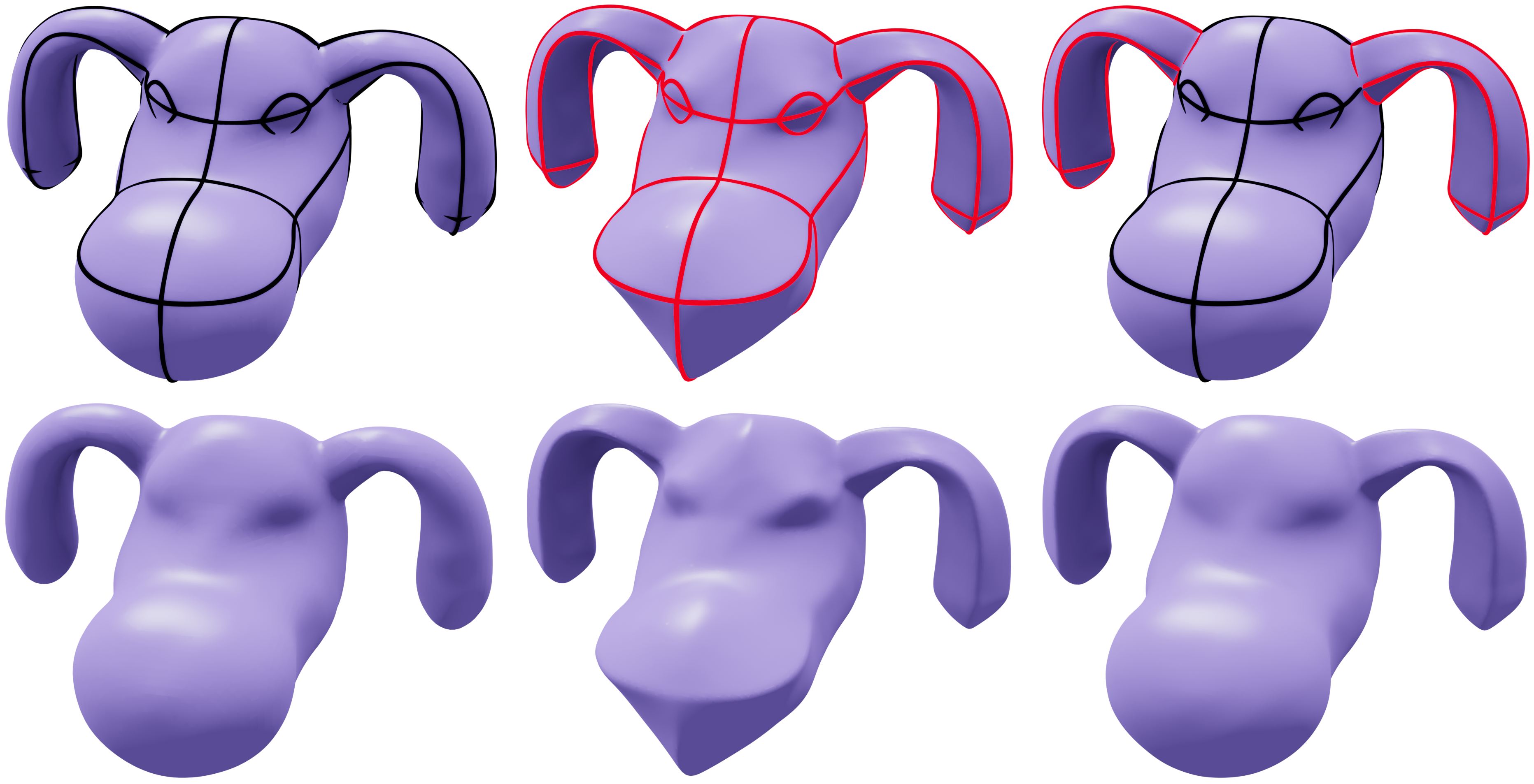}
  \makebox[0.28\columnwidth][c]{\small (a) Smoothness}
    \makebox[0.28\columnwidth][c]{\small (b) Feature}
    \makebox[0.28\columnwidth][c]{\small (c) Mixed}
  \end{center}
   \caption{Experiment on different curve types. (a) All curves are treated as smoothness; (b) all curves are treated as features; (c) specifying only the ear as a feature and incorporating it into our loss term.}
   \label{fig:abs_loss_feature}
\end{figure}
We show more results on how different types of curves influence the output. By specifying only the ear as a feature and incorporating it into our loss term, we generate a more reasonable shape that better represents the real geometry (Figure~\ref{fig:abs_loss_feature}(c)). It shows that our feature preservation strategy gives attention to both smoothness and features.

\subsection{Effect of $Q_{zero}$ Size.}
\begin{figure}[ht]
   \begin{center}
  \includegraphics[width=0.45\textwidth]{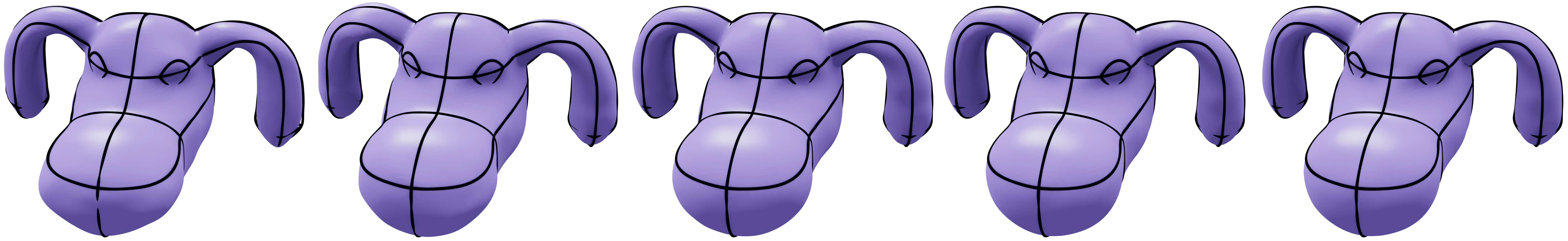}
    \makebox[0.185\columnwidth][c]{\small  (a) $N : 0.2K$}
    \makebox[0.185\columnwidth][c]{\small  (b) $N : 1K$}
    \makebox[0.185\columnwidth][c]{\small  (c) $N : 2K$}
    \makebox[0.185\columnwidth][c]{\small  (d) $N : 5K$}
    \makebox[0.185\columnwidth][c]{\small  (e) $N : 10K$}
  \end{center}
   \caption{Ablation study on sampling at the zero-level of the SDF, where $N$ is the number of sampling points. When the number of sampling points exceeds 5K, the improvements become marginal.}
   \label{fig:abl_weight_numsample}
\end{figure}

We analyze the impact of the number of $Q_{zero}$ sampling points on the construction quality. 
By adjusting the number of $Q_{zero}$ points from 0.2K to 10K, we observe that increasing the number of sampling points generally leads to more accurate surface construction, as shown in Figure~\ref{fig:abl_weight_numsample}, where $N$ denotes the number of sampling points.
However, beyond a certain threshold (e.g., 5K), the improvements become marginal, and the computational cost increases significantly. 
Therefore, a conservative choice of 10K points is adopted for all experiments, as it provides a good trade-off between accuracy and efficiency.

\subsection{Effects of Loss Function Terms.}

\begin{figure}[ht]
  \begin{center}
  \includegraphics[width=0.98\linewidth]{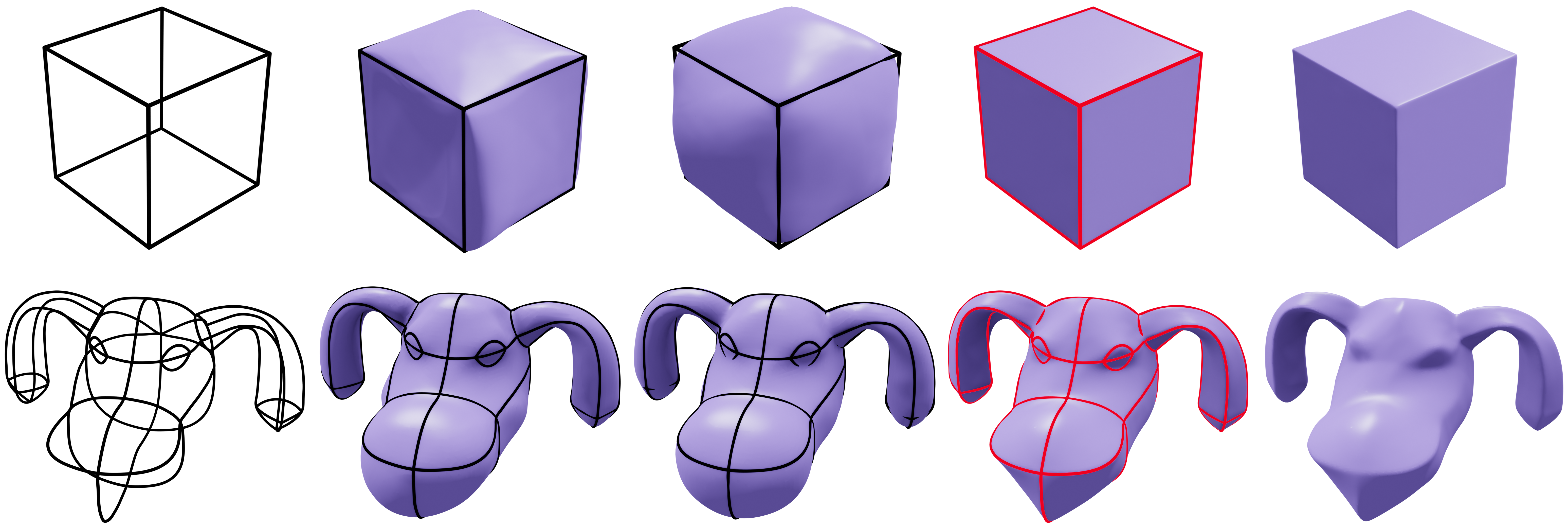}
    \makebox[0.15\columnwidth][c]{\footnotesize (a) Input}
    \makebox[0.19\columnwidth][c]{\footnotesize (b) $S(\times)$ $F(\times)$}
    \makebox[0.19\columnwidth][c]{\footnotesize (c) $S(\checkmark)$ $F(\times)$}
    \makebox[0.19\columnwidth][c]{\footnotesize (d) $S(\checkmark)$ $F(\checkmark)$}
    \makebox[0.19\columnwidth][c]{\footnotesize (e) $S(\checkmark)$ $F(\checkmark)$}
  \end{center}
   \caption{Ablation study on loss functions. $S$ represents the thin-plate energy term, and $F$ represents the feature curve (red) constraint. (b) The loss function includes only the Dirichlet and Eikonal conditions; (c) incorporating the thin-plate energy constraint; (d) a piecewise smooth surface is achieved when all curves are treated as feature curves (red).}
   \label{fig:abl_loss_func}
\end{figure}

Compared to the original version of IGR~\cite{IGR2020}, we introduce the thin-plate energy loss term to constrain the empty space between the curve networks. 
As shown in Figure~\ref{fig:abl_loss_func}, $S$ represents the thin-plate energy loss term, and $F$ denotes the feature curve constraint.
When the loss function only incorporates the Dirichlet and Eikonal conditions, it fails to produce a smooth and regular shape in the empty space between the curve networks (see Figure~\ref{fig:abl_loss_func} (b)). 
By introducing the thin-plate energy constraint, the generated shape becomes more reasonable (see Figure~\ref{fig:abl_loss_func} (c)). 
Furthermore, we obtain a piecewise smooth surface when all curves are treated as feature curves (red), as shown in Figure~\ref{fig:abl_loss_func} (d).

\subsection{Shape Evolution over Training Iterations}
\begin{figure}[ht]
  \begin{center}
  \includegraphics[width=0.45\textwidth]{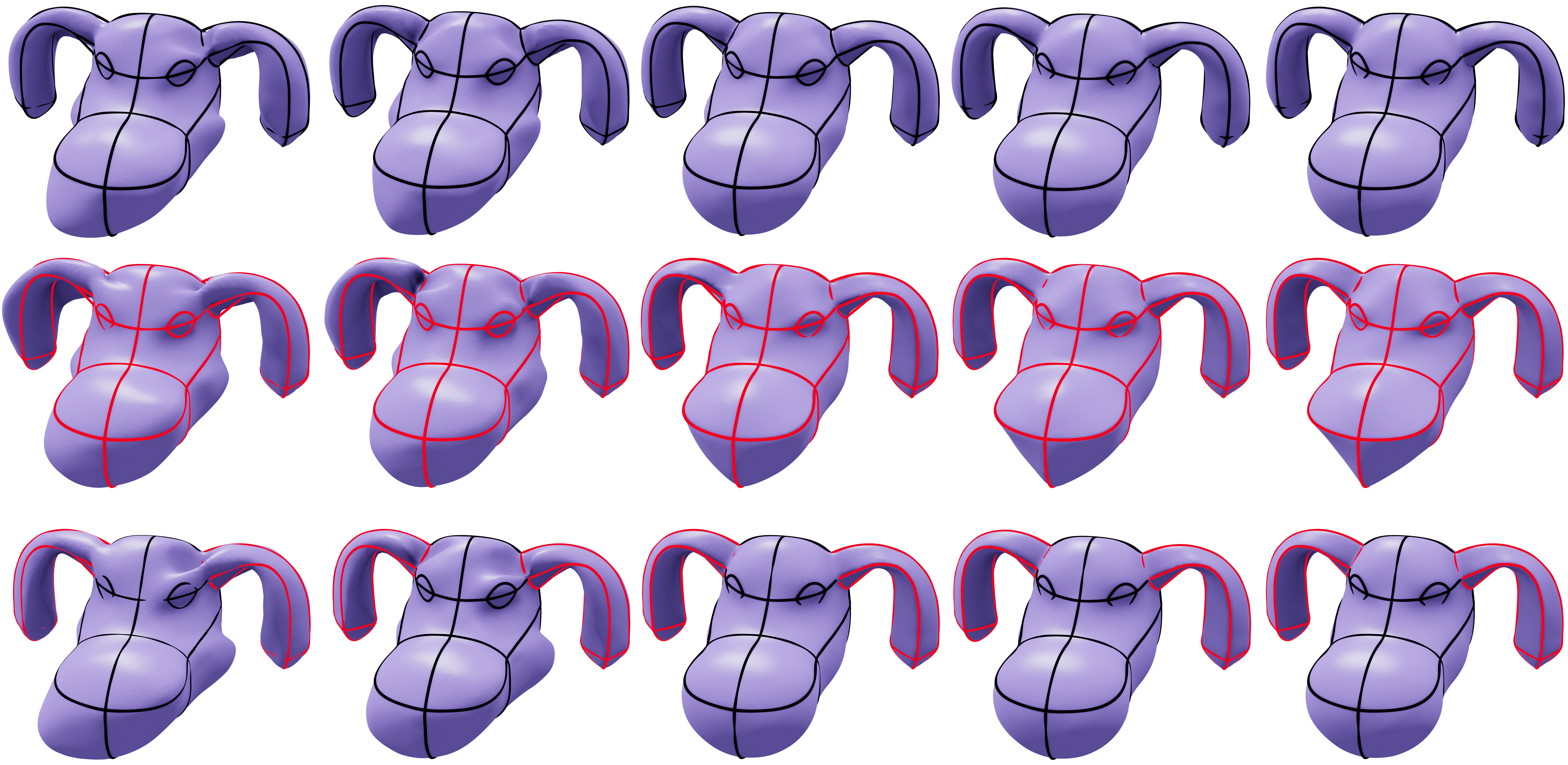}
    \makebox[0.185\columnwidth][c]{\small (a) $Iter : 1K$ }
    \makebox[0.185\columnwidth][c]{\small (b) $Iter : 3K$ }
    \makebox[0.185\columnwidth][c]{\small (c) $Iter : 5K$ }
    \makebox[0.185\columnwidth][c]{\small (d) $Iter : 7K$ }
    \makebox[0.185\columnwidth][c]{\small (e) $Iter : 9K$ }
  \end{center}
   \caption{Shape evolution over training iterations. $Iter$ represents the number of iterations. The curve types are set to smooth (top), feature (middle), and mixed (bottom), where only the ear is designated as a feature. The results show that the shape stabilizes around 5K iterations.}
   \label{fig:abl_init_iter}
\end{figure}
We conducted a series of experiments to assess the convergence speed of NeuVAS, as depicted in Figure~\ref{fig:abl_init_iter}, where $Iter$ denotes the number of iterations. 
Three different configurations of curve networks were evaluated: all curves treated as smooth, all curves as feature curves, and a mixed configuration, where only a specific part is designated as a feature. 
The results show that the shape converges after approximately 5K iterations. To ensure reliable convergence, we conservatively set the number of iterations to 10K for all subsequent experiments.

\subsection{Sharpness}
\begin{figure}[ht]
  \begin{center}
  \includegraphics[width=0.45\textwidth]{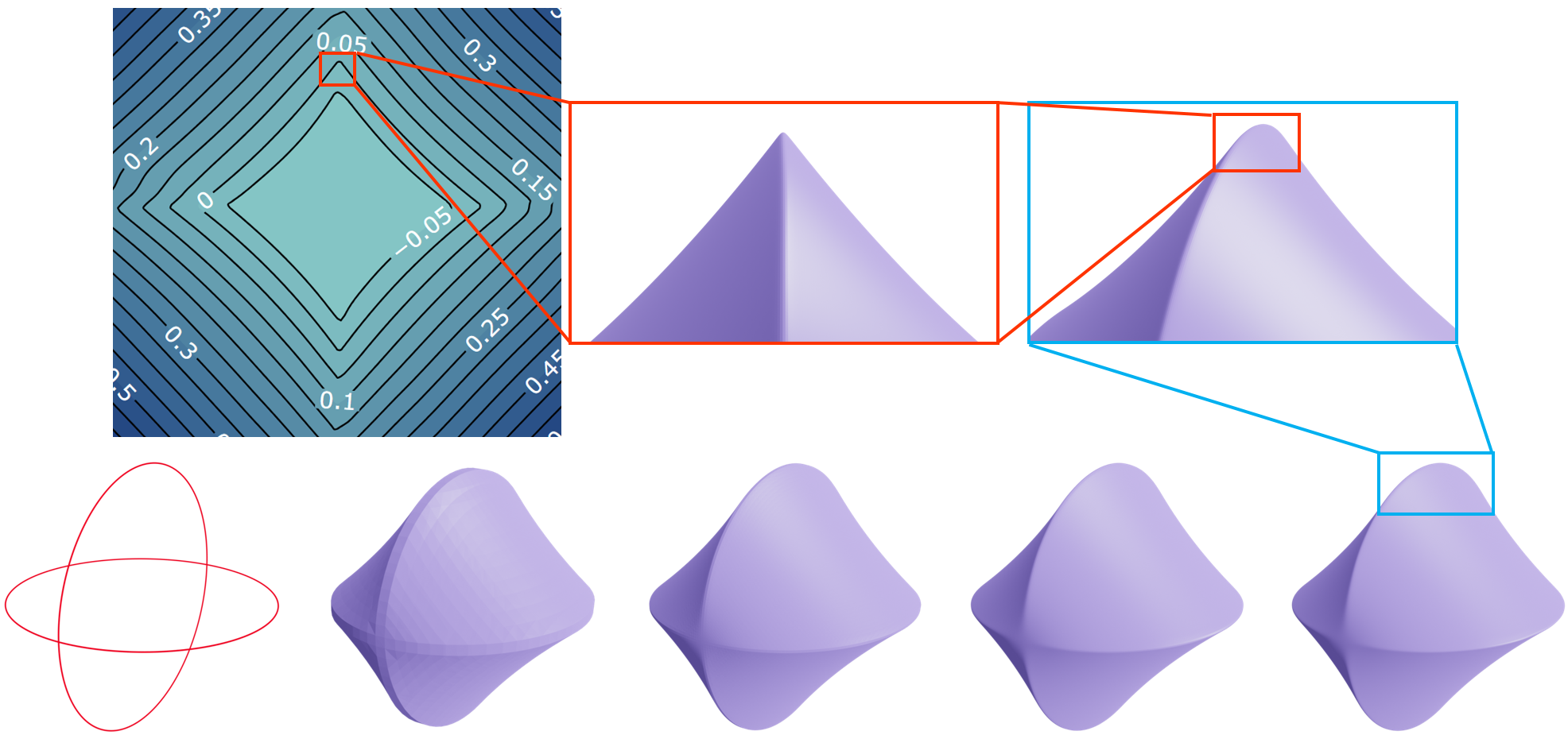}
    \makebox[0.185\columnwidth][c]{\small (a) Input }
    \makebox[0.185\columnwidth][c]{\small (b) $Res : 128^3$ }
    \makebox[0.185\columnwidth][c]{\small (c) $Res : 256^3$ }
    \makebox[0.185\columnwidth][c]{\small (d) $Res : 512^3$ }
    \makebox[0.185\columnwidth][c]{\small (e) $Res : 1024^3$ }
  \end{center}
   \caption{Marching Cubes at multiple resolutions are used to demonstrate the sharpness.}
   \label{fig:howsharpness}
\end{figure}
We employ Marching Cubes at multiple resolutions to visualize the sharpness of the constructed surfaces, as shown in Figure~\ref{fig:howsharpness}.
The resolution of the Marching Cubes is progressively increased to clearly illustrate the improvement in sharp feature, accompanied by 2D SDF slices for further clarity.
At a resolution of $1024^3$, the constructed surface closely approximates the true SDF.
Although perfect sharpness is not achieved, the results exhibit a high degree of geometric fidelity and are visually near-sharp.

\subsection{Additional datasets}
\begin{figure}[ht]
  \begin{center}
  \includegraphics[width=0.45\textwidth]{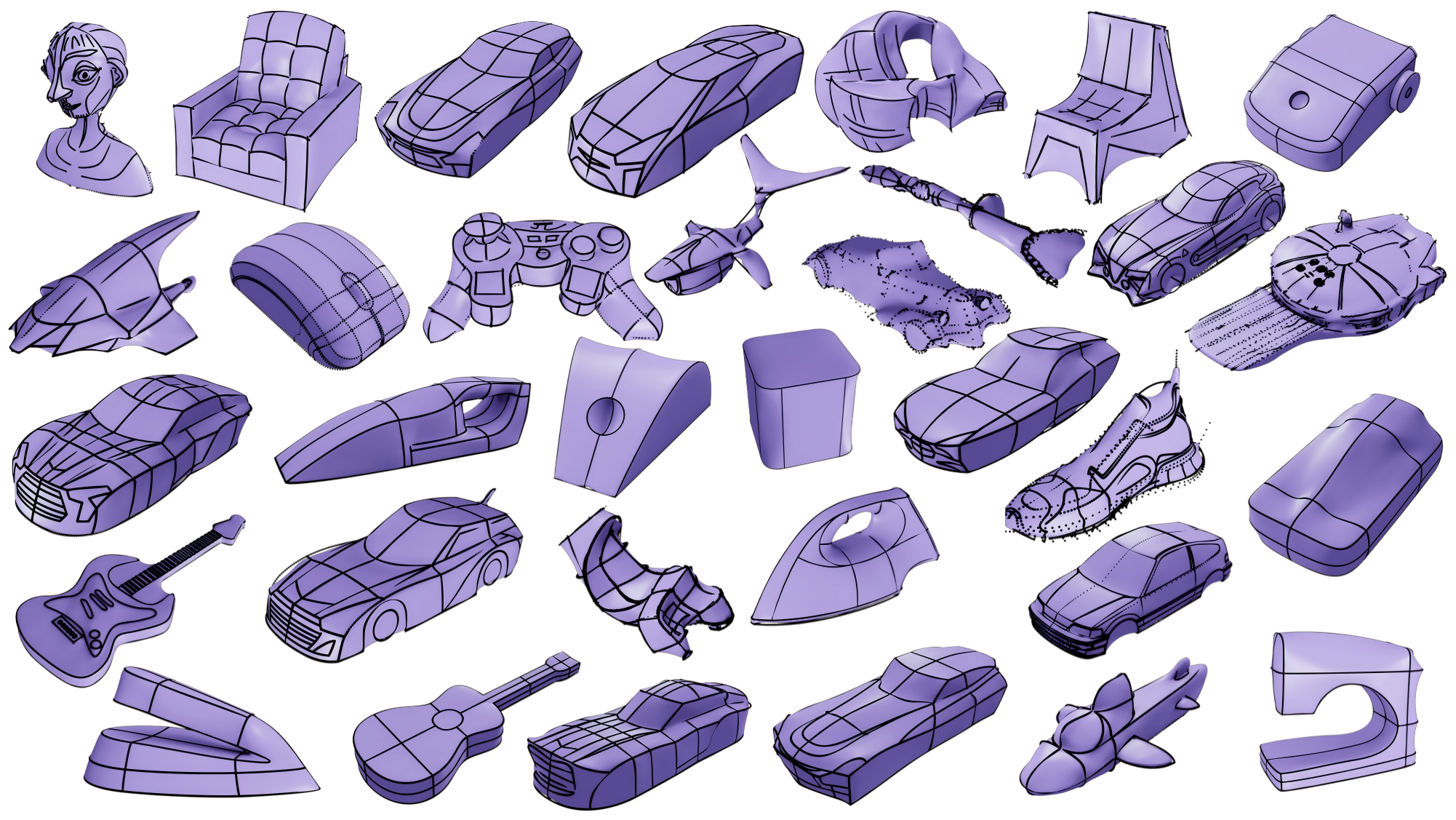}
  \end{center}
   \caption{A gallery of our results on ~\cite{yu2022piecewise} dataset.}
   \label{fig:yu2020}
\end{figure}
We present results on the ~\cite{yu2022piecewise} dataset in Figure~\ref{fig:yu2020} to further demonstrate the generality and robustness of our approach.
Since our method has difficulty handling open surfaces, models containing such geometries were excluded from the evaluation.

\end{document}